\documentclass{aa}  

\usepackage{graphicx}
\usepackage{pgfplots}
\usepackage{txfonts}
\usepackage{hyperref}
\usepackage{tabularx}
\usepackage{txfonts}
\usepackage{color}
\usepackage{xcolor}
\usepackage{natbib}
\defcitealias{<bibkey>}{<alias text>}
\defcitealias{Paxton2011}{\texttt{MESA} Paper~I}
\defcitealias{Paxton2013}{\texttt{MESA} Paper~II}
\defcitealias{Paxton2015}{\texttt{MESA} Paper~III}
\defcitealias{Paxton2018}{\texttt{MESA} Paper~IV}
\defcitealias{Paxton2019}{\texttt{MESA} Paper~V}
\defcitealias{Jermyn2023}{\texttt{MESA} Paper~VI}
\usepackage{amsmath}
\usepackage[normalem]{ulem}
\usepackage{booktabs} % Required for professional horizontal rules
\usepackage{siunitx}  % Optional, but highly recommended for formatting scientific numbers/units

\newcommand{\msun}{\text{M}_\odot}
\newcommand{\rsun}{\text{R}_\odot}
\newcommand{\mdot}{\text{M}_\odot \ \ensuremath{\text{yr}^{-1}}}
\newcommand{\yr}{\text{yr}}

\newcommand{\lsun}{\text{L}_\odot}
\newcommand{\hethree}{^{3}\text{He}}

\newcommand{\liseven}{^{7}\text{Li}}
\newcommand{\beseven}{^{7}\text{Be}}
\newcommand{\deuterium}{^{2}\text{H}}
\newcommand{\nififteen}{^{15}\text{N}}
\newcommand{\egrav}{\epsilon_\text{grav}}

\begin{document} 

%   \title{Asteroseismic imprint of accretion burst during star formation}

%   \title{Inferring past accretion bursts from asteroseismic imprints}

   \title{Inferring accretion bursts in intermediate-mass protostars from asteroseismic imprints}

%   \subtitle{Bursty burst plus seismology}

   \author{J. Jørgensen \inst{1}, 
            E. I. Vorobyov \inst{1},
            K. Zwintz \inst{1},
            E. Farag \inst{2},
            A. Carvalho \inst{3},
            and J. Henneco \inst{4}            
          }

\authorrunning{Jørgensen et al.}

   \institute{Institut für Astro- und Teilchenphysik, Universität Innsbruck, Technikerstraße 25, 6020 Innsbruck, Austria \and Department of Astronomy, Yale University, New Haven, CT 06511, USA \and Center for Astrophysics — Harvard \& Smithsonian, Cambridge, MA 02138, USA \and School of Mathematics, Statistics and Physics, Newcastle University, Newcastle upon Tyne, NE1 7RU, United Kingdom} 
   \date{Received 24 July 2026, Accepted 12 November 2026.}

% \abstract{}{}{}{}{} 
% 5 {} token are mandatory
 
  \abstract
    {During star formation, objects may gain the majority of their final mass through regular and intense bursts of mass accretion. For burst events of the FU Orionis-type (FUor), accretion rates can climb as high as $10^{-3} \ \mdot$ during short windows of time, amounting to just a few percent of the main accretion phase. Some bursters were not observed in the preceding quiescent stage, making their classification as FUors difficult. To study the occurrence of such events observationally, signatures of the current accretion bursts and tracers of recent bursts are crucial.} % leave it empty if necessary  
    {We investigate the potential to identify current accretion bursts and detect recent bursts in protostars and pre-main sequence (pre-MS) stars of intermediate mass using asteroseismology. The aim is to characterize theoretical period-spacing diagrams of pre-MS gravity-mode (\textit{g}-mode) pulsators that have undergone accretion bursts.}
    {We use the stellar structure and evolution code \texttt{MESA} to evolve stellar models from the protostellar seed phase of $0.01 \ \msun$ and $1.5 \ \rsun$ up to $4.0 \ \msun$ with a constant accretion rate of $10^{-6} \ \mdot$. A single accretion burst is initialized during the accretion phase. The stellar oscillation code \texttt{GYRE} is employed to compute the resulting asteroseismic properties of the models.}
    {Accretion bursts of up to $10^{-3} \ \mdot$ are sufficient to cause the convective cores of the models to recede, leaving behind a chemical trace which manifests itself in the form of quasi-periodic dips in the \textit{g}-mode period spacing. This effect is strongest during the outburst, but also lingers for up to 475 years before the convective core is re-established and the dips are gone.}
    {Our experiment shows that a single accretion burst leaves an imprint in the deep interiors of protostars and pre-MS stars of intermediate mass, and that this imprint can be captured with asteroseismology. Well-resolved \textit{g}-mode pulsations of young stellar objects may therefore help identify current and recent accretion bursts. Our results are most conclusive for burst magnitudes greater than $1\times10^{-4} \ \mdot$, corresponding to FUor-type bursts.}
    

   \maketitle   
   
%
%-------------------------------------------------------------------

\section{Introduction}\label{sec:introduction}

Stars form during the gravitational collapse of dense molecular cloud cores. However, most of their final mass comes not directly from the cloud but rather from a protostellar disk that forms shortly after the protostar due to non-zero angular momentum in the cloud. The character of this so-called disk-mediated protostellar accretion depends not only on the physical mechanisms that drive mass and angular momentum transport in the disk, but also on the immediate circumdisk environment, and may be unique to each object \citep{Hartmann2009}.

One plausible formation scenario is variable accretion with episodic bursts \citep{Hartmann1996,VorobyovBasu2015}. Here, the protostar acquires a substantial portion of its final mass (\citealp[up to 50\%;][]{DunhamVorobyov2012, Meyer2021}) via short but intense episodes of mass accretion, often exceeding $10^{-4} \ \mdot$ at the peak, while having low-rate quiescent accretion in between the bursts. Such bursts are known as FU Orionis-type eruptions (\citealp[FUors, after][]{1971Afz.....7..557A}) and several dozens of such objects have been identified to date \citep{Audard2014, Connelley2018, Fischer2023}. 

The exact mechanism of the burst in FUors has not been determined so far and many possibilities have been proposed in the literature (see a review in \citealp{Audard2014} and more recent mechanisms in \citealp{Dullemond2019,Borchert2022,Demidova2023,Nayakshin2023,Skliarevskii2023,Cecil2024}). Regardless of the triggering mechanism, the bursts affect the evolution of both the star and protostellar disk via energy release at the stellar surface during the burst phase, and are an important ingredient of star and planet formation. For instance, bursts can evaporate volatile species and alter the chemical composition \citep{Rab2017,Molyarova2018,Guadarrama2024}, shift the snowlines \citep{Cieza2016,Tobin2023,Lee2024}, and affect dynamics, dust growth, and refractory species in protoplanetary disks \citep{Hubbard2017,Vorobyov2020,Vorobyov2022,LiZhu2023,Lambrechts2024}.
Furthermore, bursts can affect the pre-main sequence (pre-MS) evolution of stars, contributing to a luminosity spread in stellar populations and triggering peculiar excursions on the Hertzsprung-Russell (HR) diagram in single star evolution \citep{Baraffe2012,Kunitomo2017,Jensen2018, Elbakyan2019,Meyer2019}. Bursts can also leave imprints on the stellar pulsation modes in $\delta$-Scuti and low-mass protostars \citep{Steindl2021,Steindl2022,Joerjensen2025}. 

To assess the impact of outbursts on star- and planet formation, it is important to have a census of FUors that is as complete as possible.
The FUor-type outbursts in low-mass protostars ($0.5 < M \ [\msun] < 1.5$) are short-lived events, lasting from tens to perhaps a few hundreds of years \citep{Audard2014}. Only about half of known burst sources were observed in the pre-burst quiescent phase \citep{VorobyovFUORS-2021}, making their identification as FUors firmly established. The other so-called FUor-like objects were not observed before outburst and were mainly classified as FUors based on their peculiar spectral characteristics \citep{Connelley2018}.

Furthermore, numerical models of low-mass protostar formation predict that no more than a few percent of their main accretion phase is spent in the burst phase (\citealp{DunhamVorobyov2012}, see also a review by \citealp{Audard2014}). 
Near- and mid-infrared surveys of star forming regions also suggest that quiescent accretion between the bursts can last from thousands to tens of thousands of years, much longer than the burst duration itself \citep{Scholz2013,Hillenbrand2015,Fischer2019}. Time spacings between the knots in protostellar jets imply a similar quiescent phase duration \citep{Vorobyov2018}. 

This means that low-mass protostars are difficult to catch in the outburst phase and the current census of FUors includes only a small fraction of outbursts that may have occurred during the stellar build-up. Methods to identify the past FUor bursts are based on (i) the presence of extended CO emission in the envelope that surrounds young stars in the embedded stage of their evolution \citep{Lee2007,Vorobyov2013,Freeman2017,Rab2017}, and (ii) on the evolution of dust and its spectral index in the disk during and after the burst \citep{Vorobyov2022,Houge2023}, but evaluating their efficiency requires further focused studies.

Identifying current and past FUor-type bursts in the young intermediate-mass protostars ($1.5 < M \ [\msun] < 7.0$; YIMS) may be even more challenging than in their low-mass counterparts.
Recent numerical simulations suggest that  outbursts in YIMS can be long-lived phenomena, up to several kyr \citep{Das2025}. This makes it difficult to identify the outburst objects as FUors because the onset of the burst likely occurred before photometric observations began. For example, a long-lived outburst in  RNO~54 \citep{Hillenbrand2023} was not classified as a FUor for more than half of its estimated 130+ yr outburst duration.
Spectroscopic identification as FUor-like objects in YIMS is also associated with difficulties as discussed in Sect.~\ref{sec:discussion}.
Therefore, having a definitive identifier of a source that is undergoing a FUor outburst, particularly in the rarer YIMSs, will help improve the completeness of the known sample of FUors.

Here, we propose a method to potentially identify current and recent episodic luminosity bursts by probing their impact on the protostar's asteroseismic properties. Asteroseismology—the study of stellar oscillations and their link to stellar interiors \citep[see, for example,][]{Aerts2021}—has shown that stars across many evolutionary classes can become unstable to pulsations. During the pre-MS phase, models that grow from $0.01\ \msun$ seeds via bursty accretion exhibit distinct asteroseismic signatures compared to those evolved using classical pre-MS prescriptions, where models are initialized already at their final masses, and contract along the Hayashi and Henyey tracks \citep{Henyey1955, Hayashi1961, Iben1965}. Specifically, pressure-mode oscillation frequencies in $\delta$-Scuti and low-mass stars show noticeable differences depending on the adopted accretion history \citep{Steindl2022,Joerjensen2025}. 

In this study, we instead turn our attention to the family of pulsation modes named gravity modes (\textit{g}-modes), for which the restoring force is buoyancy, and which are sensitive to, among other things, the near-convective-core boundary region \citep{Pedersen2018}. We specifically target YIMSs, as they are predicted to become unstable against \textit{g}-mode  pulsations on their approach towards the zero-age main-sequence \citep[ZAMS;][]{Steindl2021}, and pulsations are observationally confirmed for these objects, both during the pre-MS and the early ZAMS \citep{Zwintz2013-gdor,Zwintz2017-SPB}. We employ release 25.12.1 of the \texttt{MESA} stellar evolution code \citep[Modules for Experiments in Stellar Astrophysics, Papers I-VI;][]{Paxton2011, Paxton2013, Paxton2015, Paxton2018, Paxton2019, Jermyn2023} and the \texttt{GYRE} stellar oscillation code \citep{GYRE2013, GYRE2018, GYRE2020, GYRE2023} to study the immediate response of the stellar structure and, in particular, its \textit{g}-mode pulsations to a mass accretion burst.  

The paper is organized as follows: In Sect. \ref{sec:MESA} we summarize our numerical setup for computing pre-MS models with accretion bursts in \texttt{MESA}, as well as our setup for \texttt{GYRE}. Sect. \ref{sec:Results} covers our results for the stellar evolution and asteroseismic properties of FUors. We discuss these results and put them into perspective in Sect. \ref{sec:discussion}. Lastly, we summarize this work and draw our conclusions in Sect. \ref{sec:conclusion}.
%--------------------------------------------------------------------
\section{Numerical model}\label{sec:MESA}

We begin by outlining our \texttt{MESA} setup for calculating a self-consistent pre-MS evolution with a constant accretion rate. We then describe the implementation of a single accretion burst. Lastly, we explain our \texttt{GYRE} setup.

\subsection{Accretion in \texttt{MESA}}\label{sec:accretion_in_mesa}
\texttt{MESA} is a Henyey-style stellar evolution code \citep{Henyey1963} that solves the coupled equations of stellar structure and evolution simultaneously on a one-dimensional Lagrangian mesh. The code includes a built-in prescription for mass accretion, which can be activated by specifying a mass accretion rate via \texttt{mass\_change} in units of solar masses per year (or equivalently \texttt{mdot\_star} in cgs units). One aspect of this work is to analyze the physical response of our stellar models in the face of accretion. We therefore provide a brief overview of how \texttt{MESA} treats accretion. 

Accretion was introduced in \citetalias{Paxton2011}: with a specified mass accretion rate $\dot{M}$, a mass $\delta M=\dot{M}\delta t$ is added over a timestep $\delta t$. Accretion delivers new material to the surface, hence \texttt{MESA} solves for both the new structure and composition. To accommodate the added mass, a model is partitioned into two regions \citepalias[Figure 32 of][]{Paxton2015}: an inner Lagrangian region where cells remain constant in mass coordinate \textit{m}, and an outer homologous region, in which cells remain constant in relative mass coordinate $q=m/M$. An intermediate region also exists; it is calculated as a weighted average of the two. This partitioning is preferable when the thermal profile can vary rapidly \citep{Hayashi_Sugimoto1965, Sugimoto1975, Sugimoto1977} and allows \texttt{MESA} to handle the rapid accretion rates explored in this work \citepalias{Paxton2015}.

The local response to accretion can be understood in terms of the quantity $\egrav$, which emerges from an imbalance in the energy equation:
\begin{equation}\label{eq:energy_equation_with_eps_grav_and_dsdt}
    \frac{\partial L}{\partial m}-\epsilon_\text{nuc}=\egrav=-T\left(\frac{\partial s}{\partial t}\right)_m,
\end{equation}
where $\partial L / \partial m$ is the luminosity gradient, $\epsilon_\text{nuc}$ is the net nuclear energy generation rate, including neutrino losses, $T$ is temperature, and $(\partial s / \partial t)_m$ is the rate of change of specific entropy.
$\egrav$ therefore represents the rate of heat added or removed from a cell ($dQ/dt$; \citealp[chapter 6.1 in][]{HansenBook2004}). From thermodynamic relations, we find that such changes are driven simultaneously by changes in internal energy ($du/dt$) and work done to either expand or compress the stellar material locally ($P \ d(1/\rho)/dt$):
\begin{equation}\label{eq:egrav_sect_2}
    \egrav=-\frac{du}{dt}-P\frac{d}{dt}\left(\frac{1}{\rho}\right)=-T\left(\frac{\partial s}{\partial t}\right)_m.
\end{equation}
Therefore, $\egrav$ may take on positive or negative values, representing heat added or removed---usually corresponding to either contraction or expansion---depending on the balance between the rate of internal energy and work. However, Eq. (\ref{eq:egrav_sect_2}) also reveals that the local response may be interpreted in terms of entropy; $(\partial s/\partial t)_m=-\egrav/T$. Regions with $(\partial s/\partial t)_m>0$ are losing heat, whereas regions with $(\partial s/\partial t)_m<0$ are gaining heat. In general, heat is lost because the star must accommodate the added entropy, usually by expanding.

A complementary view on the preceding discussion considers timescales (\citetalias{Paxton2015, Paxton2018, Paxton2019}; \citealp{Hamstars}). During the pre-MS, nuclear burning does not set the evolutionary timescale. Instead, the star evolves on a thermal timescale set by the ratio of total energy to the radiant luminosity. This so-called Kelvin-Helmholtz timescale can be written as
\begin{equation}
    \tau_\text{KH} = \frac{GM^2}{RL},
\end{equation}
where $G$ is the Newtonian gravitational constant, $M$ is the total stellar mass, $R$ is the total stellar radius, and $L$ is the luminosity at the surface \citep{KippenhahnBook2013}. Accretion introduces the accretion time-scale,
\begin{equation}
    \tau_{\dot{M}}=\frac{M}{\dot{M}},
\end{equation}
where $\dot{M}$ is the mass accretion rate. When $\tau_{\dot{M}}\ll\tau_\text{KH}$ the star cannot freely radiate away the added heat, leading to entropy accumulation in the outer layers and radial inflation. Conversely, for $\tau_{\dot{M}}\gg\tau_\text{KH}$, the star can thermally relax and evolves on $\tau_\text{KH}$. 

\subsection{Stellar models}\label{subsec:stellar_models}
The stellar models calculated with \texttt{MESA} are tailored for both the pre-MS evolution, accretion, and asteroseismology. We summarize here some important input physics and \texttt{MESA} settings in the models.These considerations will be important when evaluating the response of the star in the face of an accretion burst. Notably, our choices for values of input parameters, for example, related to mixing or chemical composition, are mostly fiducial as we are neither calibrating our models to observations nor exploring a parameter space. For additional information on our chosen input physics and settings we refer to Github\footnote{\url{https://github.com/Johanneshj/Asteroseismic-imprint-of-accretion-burst}} where the full setup is available. The setup is also published on Zenodo\footnote{\url{https://doi.org/10.5281/zenodo.24249376}} along with the data to reproduce our results.

For high temperatures we adopt the OPAL opacity tables \citep{Iglesias1993, Iglesias1996}, while the tables of \cite{Ferguson2005} are used for low-temperatures. We also use the proto-solar abundances of \cite{Asplund2009}. The default \texttt{MESA} equation of state is a blend of tables from OPAL/SCVH \citep{Rogers2002, Saumon1995}, FreeEOS \citep{Irwin2012}, HELM \citep{Timmes2000}, and SKYE \citep{Jermyn2021}.

For the treatment of convection, the options in \texttt{MESA} broadly follow a mixing-length theory in which fluid parcels are assumed to traverse a characteristic length---the so-called mixing-length $\alpha_\text{mlt}$---in the stellar interior before dispersing and releasing their retained heat to their surroundings \citep{MJoyce2023}. We choose the option \texttt{Cox} \citep{Cox&Guili1968} which is an extension of the canonical mixing-length theory by \cite{BohmVitense1958} and set $\alpha_\text{mlt}=2.0$. We verified that our results are insensitive to the choice between the \texttt{Cox} and \texttt{Henyey} formulations, appropriate for optically thick and thin regimes respectively, and which differ with the inclusion of radiative losses in the \texttt{Henyey} formulation. 

% Time-dependent convection (\texttt{TDC}; \citealp{Jermyn2023}) is not considered, although it may become relevant given the short timesteps in some phases. 

To determine which zones are unstable against convection we use the Ledoux criterion, where the radiative gradient ($\nabla_\text{rad}$) is compared against the Ledoux gradient ($\nabla_\text{L}=\nabla_\text{ad}+\nabla_\mu$), with convective instability at $\nabla_\text{rad}>\nabla_\text{ad}+\nabla_\mu$ \citep{Ledoux1947}. In regions unstable according to the Schwarzschild criteria (\citealp[$\nabla_\text{rad}>\nabla_\text{ad}$][]{Schwarzschild1958}), but stable against Ledoux, we use semiconvection following the prescription of \cite{Langer1983} with an efficiency $\alpha_\text{sc}=0.1$. For convective boundary mixing we turn on exponential overshooting according to \cite{Herwig2000} with over- and undershooting parameters $f/f_0=0.005/0.001$ and $f/f_0=0.0025/0.001$ for top and bottom of convection zones respectively. Here $f_0$ represents the distance into the convection zone from where the convective diffusion coefficient, which serves as the maximum diffusion coefficient of the exponentially decreasing convective-boundary mixing profile, is taken, and $f$ is the overshooting distance, both in units of the pressure scale height. As per recommendation from \cite{Buchele2025}, we also set the minimum diffusive mixing coefficient $D_\text{mix} = 0.01 \ \text{cm} \ \text{s}^{-1}$. This setting was discussed in \cite{Wagg2024} who noted fewer numerical glitches in key quantities, such as the Brunt-Väisälä frequency and the mean molecular weight, for larger values of $D_\text{mix}$ and found a compromise at $D_\text{mix}=20 \ \text{cm} \ \text{s}^{-1}$ (\citealp[see also][]{Farag2022:BH}). We similarly tested values from $0.01$ to $100 \ \text{cm} \ \text{s}^{-1}$, but found no significant differences between Brunt-Väisälä profiles.

For the chemical composition we choose an initial metallicity of $Z=0.02$ and obtain the initial helium abundance ($Y$) from a galactic chemical enrichment law $\Delta Y/\Delta Z = 1.4$ \citep{Ricci2025}. We additionally specify the initial abundance-ratios of deuterium ($^2\text{H}/^1\text{H}$) and helion ($^3\text{He}/^4\text{He}$); these are set to 20 ppm and 85 ppm respectively \citep{Gloeckler1998,Mahaffy1998,vandevoort2018,Steindl2021}. The initial hydrogen abundance is then obtained as $X=1-Y-Z-X_{^2\text{H}}-X_{^3\text{He}}$. The initial chemical composition also represents the abundances present in the primordial birth cloud. We therefore instruct \texttt{MESA} to accrete material with this initial composition onto the surface (toggled with \texttt{accrete\_same\_as\_surface = .false.}).

The atmosphere represents the outer boundary condition of the stellar models and similarly the pulsation calculations with \texttt{GYRE}. We adopt an Eddington $T(\tau)$ atmosphere in which the equation of hydrostatic equilibrium is integrated incrementally at each level of optical depth to produce the temperature, pressure, and density profiles throughout the atmosphere \citep{Eddington1926}. 

The convective core boundary is also important as it acts as a turning point for the internal gravity waves (\citealp[see, for example,][]{Miglio2008, AertsBook2010, Pedersen2018}). We ensure sufficient resolution here by utilizing \texttt{MESA}'s adaptive mesh techniques for steep gradients in element species. We also set a minimum change in cell mass of $10^{-4}$ for composition changes in both radiation- and convection zones (\texttt{min\_dq\_for\_xa} in \texttt{MESA}; default value is $10^{-14}$). For values lower than this our models were not smooth in composition, and the convective-core boundary---determined from the Ledoux criterion---could not robustly be identified.

Finally, we do not assume full hydrostatic equilibrium and allow our models to evolve with non-zero cell-face velocities (\texttt{v\_flag} option in \texttt{MESA}). We adopt the \texttt{dedt} form of the energy equation, which improves numerical energy conservation \citepalias{Paxton2011, Paxton2013, Paxton2018}. Returning to the discussion of $\egrav$ in Sect. \ref{sec:accretion_in_mesa}, we note that in this formulation $\egrav$ does not explicitly appear. Instead, its effects are incorporated into the total energy evolution. We can, however, evaluate $\egrav$ by directly computing it as,
\begin{equation}\label{eq:egrav_paxton_2018}
    \egrav=-\left(\frac{De}{Dt}+P\frac{\partial}{\partial m}(vA)\right),
\end{equation}
where $De/Dt$ is the total derivative of specific energy, \textit{P} the local pressure, \textit{v} the cell velocity, and \textit{A} the cell-face area \citepalias{Paxton2018}. Both terms exists as \texttt{MESA} outputs; $De/Dt$ is obtained from \texttt{dE} divided \texttt{log\_dt\_sec}, and $P\partial (vA)/\partial m$ from \texttt{dwork\_dm}. The \texttt{dedt} formulation also requires an additional source term, $\epsilon_{\dot{M}}$ \citepalias{Paxton2019}. 
This term accounts for the fact that, during the mass-adjustment step, material entering or leaving a cell carries its own specific energy, thereby modifying the model’s energy content on a cell-by-cell basis. When $\egrav$ is explicitly computed from thermodynamic quantities \citepalias[Eq. 12;][]{Paxton2011} this corrective term is not needed. Numerical convergence issues can arise with the inclusion of $\epsilon_{\dot{M}}$, specifically regarding energy conservation \citep{WongBildsten2023,Joerjensen2025}. Our models, however, are robust against its inclusion, requiring very few timestep retries if none at all.

\subsection{Pre-MS evolution with constant accretion rate}\label{subsec:constant_rate}
Our \texttt{MESA} setup for the constant-accretion scenario resembles that of \cite{Steindl2021} (\citealp[see also][]{Steindl2022II, Steindl2022} which in turn follows earlier work:
\citealp{Stahler1980I, Stahler1980II, Siess1997, Baraffe2009, Hosokawa2011, Baraffe2012, Kunitomo2017, Jensen2018, Elbakyan2019}). 
The calculations are initialized from a protostellar seed of $0.01 \ \text{M}_\odot$ and $1.5 \ \text{R}_\odot$, corresponding to a second Larson's core \citep{Larson1969}, which is constructed with \texttt{MESA}'s \texttt{relax\_mass\_scale} option. 
For this initial model the age is set to zero. Mass accretion is then switched on: during the first 4000 years we allow the accretion rate to gradually increase from $10^{-12} \ \mdot$ up to a constant rate $\dot{M}_\text{ini}$. This gradual increase is helpful in easing the model into accretion and we describe it with the following sigmoid function:
\begin{equation}
    \begin{split}
    \dot{M} = 10^{-12} \ \mdot + \left(\dot{M}_\text{ini}- 10^{-12} \ \mdot\right) ... & \\ \times \left[1+\exp{\left(-\frac{1}{100}\left(t-\frac{4000}{2}\right)\right)}\right]^{-1},
    \end{split}
\end{equation}
where $t$ is age in years. In this work we consider only $\dot{M}_\text{ini}=10^{-6} \ \mdot$; a representative time-averaged value for the formation of intermediate-mass stars, but by no means the only rate they can attain \citep{Steindl2022II, Indrani2025}. With the values stated above we accrete in total $0.002 \ \msun$ during the ramp-up, meaning that the mass of the protostellar seed afterwards is $0.012 \ \msun$.

From here the model is evolved at the constant rate of $\dot{M}_\text{ini}=10^{-6} \ \mdot$ until we are within $0.05 \ \msun$ of the mass the star would have at the zero-age main-sequence ($M_\text{ZAMS}$; the quantity that one would usually denote the initial mass). We then gradually decrease the accretion rate from $\dot{M}_\text{ini}$ to a lower bound, $10^{-9} \ \mdot$, over 500,000 years. This ramp-down is an exponential decrease with a characteristic timescale $\tau$:
\begin{equation}
    \dot{M} = 10^{-9} \ \mdot + \left(\dot{M}_\text{ini} - 10^{-9} \ \mdot \right)\exp{\left(-\frac{t-t_\text{dec}}{\tau}\right)},
\end{equation}
where $t_\text{dec}$ is age of the beginning of the decrease in years. This age follows directly from the known initial conditions which are the starting mass and age ($0.012 \ \msun$, 4000 years), the accretion rate ($10^{-6} \ \mdot$), and the mass at the decrease ($M_\text{ZAMS}-0.05 \ \msun$). For a $4.0 \ \msun$ model the age at the decrease ($t_\text{dec}$) is therefore,
\begin{equation}
    t_\text{dec}=4000 \ \yr + \frac{(4.0-0.05-0.012) \ \msun}{10^{-6} \ \mdot}=3.942.000 \ \text{years}.
\end{equation}
We determine $\tau$ such that the integrated mass over 500,000 years equals $0.05 \ \msun$. For the values stated above, mainly $\dot{M}_\text{ini}=10^{-6} \ \mdot$, we obtain $\tau\simeq 49551.6$ years. The implementation of the ramp-down is motivated mainly by the idea accretion should not end abruptly when $M_\text{ZAMS}$ is reached \citep[see, for example, the accretion histories of][]{VorobyovBasu2015}. Generalizing this ramp-down for arbitrary values of $\dot{M}_\text{ini}$, or potentially adopting an analytical model \citep{Redkin2025}, will be the subject of future work. Once $M_\text{ZAMS}$ is reached the accretion rate drops to zero.

\subsection{Accretion burst}\label{subsec:accretion_burst}
We describe here the implementation of a single accretion burst in \texttt{MESA}. FUors are characterized by a variety of brightness curves \citep{Connelley2018,Semkov2021}, which may reflect the corresponding variations in the underlying burst-triggering mechanisms and disk physical conditions \citep{Audard2014}. Nevertheless, characteristic features for the majority of FUors, and in particular for those induced by the magneto-rotational instability \citep[MRI;][]{Bae2014,VorobyovKhaibrakhmanov2020},  can be summarized as follows: a sharp rise to a maximum brightness on timescales of a few years, an active phase lasting from tens to a few hundreds of years, followed by a slow decline to a pre-burst stage. 

\begin{figure}[h!]
    \centering
    \includegraphics[width=1.0\linewidth]{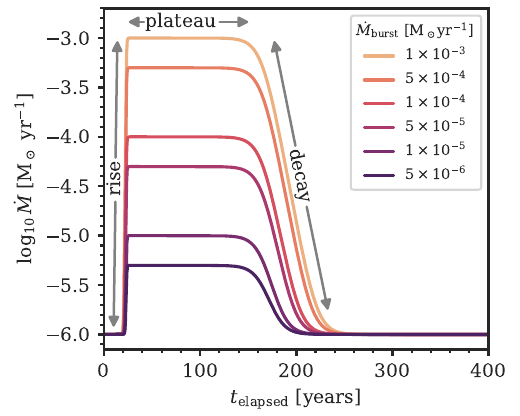}
    \caption{Burst profiles following Eq. (\ref{eq:burst_profile}) with parameters as described in the text.}
    \label{fig:burst_profile}
\end{figure}

We therefore use a toy model of the burst, which shares these characteristic features and define
the mass accretion rate by the following analytical expression:
\begin{equation}\label{eq:burst_profile}
    \dot{M}_\text{burst} = \dot{M}_\text{ini} 
    + (\dot{M}_\text{burst} - \dot{M}_\text{ini}) \left[\sigma_\text{rise}(t_\text{elapsed})-\sigma_\text{decay}(t_\text{elapsed})\right],
\end{equation}
where $\sigma_\text{rise}$ and $\sigma_\text{decay}$ are sigmoid functions defined as,
\begin{equation}\label{eq:burst_profile_sigmoids}
    \begin{split}
        & \sigma_\text{rise}(t) = \left[
            1 + \exp\!\left(
            \frac{t - \tfrac{1}{2}t_\text{rise}}{t_\text{rise}}
            \right)
            \right]^{-1}, \\
        & \sigma_\text{decay}(t) = \left[
            1 + \exp\!\left(
            \frac{t - (t_\text{rise} + t_\text{plateau} + \tfrac{1}{2}t_\text{decay}}{t_\text{decay}}
            \right)
            \right]^{-1}.
    \end{split}
\end{equation}
The resulting burst profile is a double sigmoid spanning a sharp rise ($t_{\text{rise}}$), a plateau ($t_{\text{plateau}}$), and a slow decay ($t_{\text{decay}}$). The accretion rate outside the burst is $\dot{M}_\text{ini}$, and during the burst the rate reaches a maximum $\dot{M}_\text{burst}$. Time elapsed since the onset of the burst ($t_\text{elapsed}$) is denoted \textit{t}. Fig. \ref{fig:burst_profile} illustrates all burst profiles considered in this work. Burst magnitudes ($\dot{M}_\text{burst}$) range from $5\times10^{-6}$ to $10^{-3} \ \mdot$, and remaining parameters take values $\dot{M}_\text{ini}=10^{-6} \ \text{M}_\odot \ \text{yr}^{-1}$, $t_\text{rise}=4 \ \text{yr}$, $t_\text{plateau}=100 \ \text{yr}$, and $t_\text{decay}=80 \ \text{yr}$. Importantly, \cite{Das2025} demonstrate that $t_\text{plateau}$ may be up to 500 years in the formation of intermediate stars. We intentionally choose a smaller value of 100 years to be on the conservative side. In Fig. \ref{fig:burst_profile}, the rise, plateau, and decay phases are indicated with arrows.  

To incorporate the burst we utilize the \texttt{other\_adjust\_mdot} routine in \texttt{MESA} (also used for the initial increase of the accretion rate, and the final ramp-down in Sect. \ref{subsec:constant_rate}). This routine allows for the code to dynamically update the accretion rate, meaning we are not limited to a constant accretion scenario only. Our routine requires the burst shape specified (Eq. \ref{eq:burst_profile}) and also a target mass for when to initiate the burst. At this target mass, the accretion rate departs from the constant rate in favor of the specified burst profile. Because the burst is relatively short ($\sim$100 years) compared to usual timestep sizes ($\delta t\simeq1000-10000 \ \yr$), the model would typically evolve past the burst without resolving it. We therefore first evolve the model up to an age corresponding to roughly 10 years before it reaches the target mass. In case \texttt{MESA} overshoots this age a retry is demanded with a shorter timestep. This is repeated until we reach the target age within a mass-threshold of $10^{-6} \ \msun$. The calculations are then terminated and a \texttt{pre\_burst\_model} is saved. We then initialize a new \texttt{MESA} run from the saved model with a lower timestep of $\delta t=2 \ \text{weeks}$. This is sufficient in order for the code to catch the burst, and from here the timestep sizes are handled internally by imposing a limit of $0.01 \ \mdot$ on the allowed change in accretion rate per timestep. We terminate this evolutionary run 1000 years after the accretion rate has returned to the constant rate and save the resulting model as a \texttt{post\_burst\_model}. To complete the pre-MS evolution, we similarly restart the calculations from this model and evolve it until the central hydrogen abundance drops below 0.5, terminating it during the main-sequence.

\subsection{\texttt{GYRE} setup}
We employ version 7.2.1 of \texttt{GYRE}, which solves the fourth-order boundary-value problem defined by the dimensionless and adiabatic stellar oscillation equations. We use a sixth-order Magnus multiple shooting scheme (\texttt{MAGNUS\_GL6}) to solve these equations. Outer boundary conditions are set following \citet{Dziembowski1971}, consistent with a $T(\tau)$ atmosphere, while regularity is imposed at the inner boundary such that displacements and gravitational potential perturbations vanish. Frequencies of adiabatic dipole \textit{g}-modes are computed on an inverse frequency grid with 2000 points spanning 0.05 $\text{days}^{-1}$ to 10 $\text{days}^{-1}$. We restrict solutions to modes of monotonically (negatively) increasing mode-order following the Eckart-Scuflaire-Osaki-Takata scheme \citep{Takata2006}, covering both low- and high-order \textit{g}-modes from $n_\text{pg}=-1$ to $-50$. For selected models, we compared adiabatic and non-adiabatic frequencies and found no significant differences.

\section{Results}\label{sec:Results}
We proceed now by showcasing the accreting pre-MS evolution calculated with \texttt{MESA}. For the following analysis we evolve models with ZAMS masses of $4.0 \ \msun$ and initialize an accretion burst at $3.5 \ \msun$ which, in general, is deemed the low end of the mass range of the so-called slowly-pulsating B-type stars (\citealp[SPB stars;][]{AertsBook2010}). These are generally main-sequence stars with masses 3.5 to 8 $\msun$ and spectral types B3-B8. They show multi-periodic patterns of high-order \textit{g}-modes excited by the metal opacity bump located at $T\sim2\times10^5 \ \text{K}$ \citep{Dziembowski1993} with periods between 0.5 and 5 days. During the pre-MS phase of evolution, stars at these intermediate masses may also become unstable against \textit{g}-mode pulsations, denoting a class of pre-MS SPB stars \citep{Zwintz2017-SPB,Steindl2021,Zwintz2022}. We exploit this result to examine the period-spacing patterns of our models undergoing accretion bursts, arguing that they are representative of pre-MS SPB stars. The implemented burst profiles have the following characteristics: $t_\text{rise}=4 \ \yr$, $t_\text{plateau}=100 \ \yr$, $t_\text{decay}=80 \ \yr$, and $\dot{M}_\text{burst}$ values from $5\times10^{-6}$ to $10^{-3} \ \mdot$ (see Fig. \ref{fig:burst_profile}). We highlight both the excursions in the HR-diagram caused by the accretion bursts, and also the internal structures of the models.

\subsection{Evolutionary tracks in the HR-diagram}
\begin{figure*}[t]
    \includegraphics[width=1.0\linewidth]{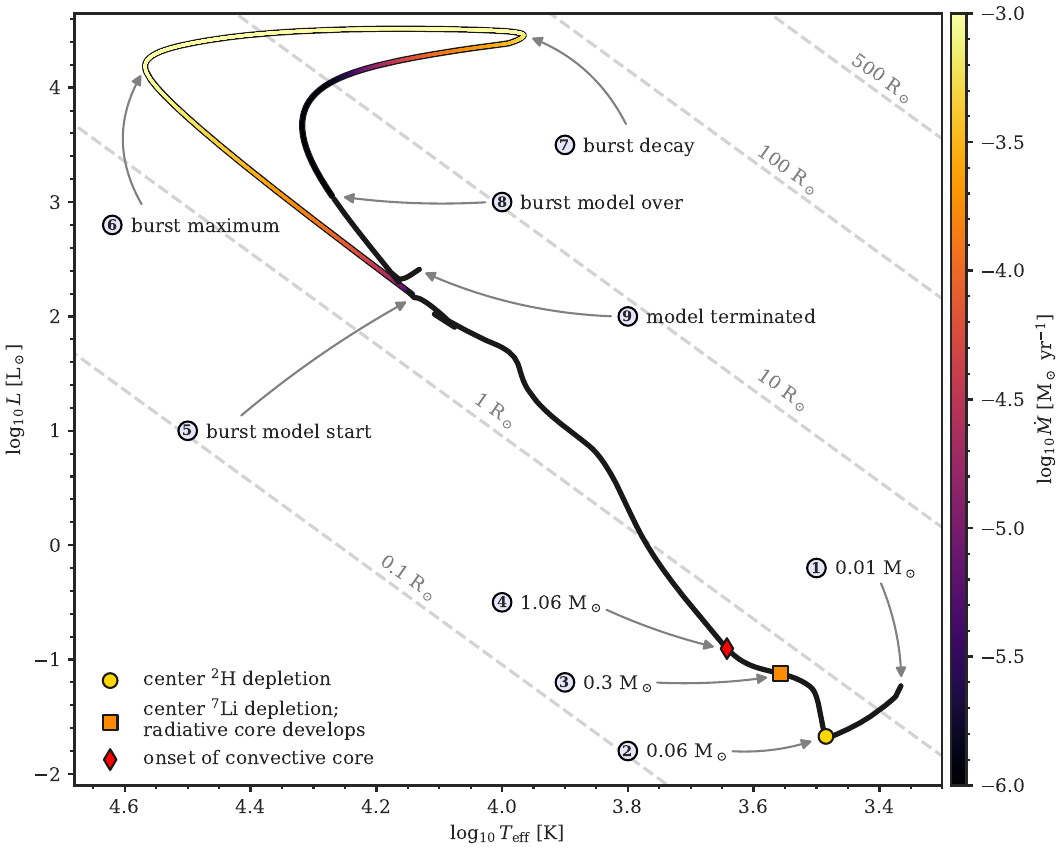}
    \caption{HR-diagram illustrating the evolutionary track of a model that undergoes an accretion burst of $\dot{M}_\text{burst}=10^{-3} \ \text{M}_\odot \ \text{yr}^{-1}$. The evolutionary track is colored by $\dot{M}$, increasing from black to yellow. Key windows during the evolution are highlighted: 1. Initial seed phase, 2. $\deuterium$ ignition, 3. $\liseven$ depletion and development of radiative core, 4. Convective core starts growing, 5. End of \texttt{pre\_burst\_model}; beginning of \texttt{burst\_model}, 6. $\dot{M}$ reaches $10^{-3} \ \mdot$, 7. $\dot{M}$ decreases again, 8. End of \texttt{burst\_model}; \texttt{post\_burst\_model} begins, 9. Central $^1\text{H}$ reaches 0.5; model is terminated.} 
    \label{fig:HR-diagram}
\end{figure*}

We draw in Fig. \ref{fig:HR-diagram} the full evolutionary track for the model undergoing an accretion burst of $10^{-3} \ \mdot$. The evolutionary track is colored by the magnitude of the accretion rate, increasing from black to yellow, emphasizing the burst excursion. Lines of constant radius are also drawn (dashed gray). Arrows guide the eye and highlight key windows during the pre-MS evolution which we now dissect. 

The track originates from the bottom right corner of the HR-diagram where we find the protostellar seed of $0.01 \ \msun$ and $1.5 \ \rsun$ (point 1). Initially, the model contracts and heats up until, at $0.06 \ \msun$, the central temperature is sufficiently hot ($\log_{10}{T_\text{c}/\text{K}}\simeq5.95$) to ignite deuterium (point 2). This is followed by an almost vertical trajectory in the HR-diagram as the star expands and gets brighter. Around $0.3 \ \msun$ central $^7\text{Li}$ is gradually depleted as the convective envelope recedes and a radiative core develops (point 3). Later, at $1.06 \ \msun$, with a central temperature of $\log_{10}{T_\text{c}/\text{K}}\simeq7.14$, a convective core starts growing when the first steps in the Carbon-Nitrogen-Oxygen (CNO) cycle can proceed, causing a buildup of $^{13}\text{C}$ (point 4). An accretion burst is initialized at $3.5 \ \msun$ (point 5). As the accretion rate grows the luminosity increases by more than two orders of magnitude, from roughly $2.0$ to $4.2$ in $\log_{10}{L/\lsun}$. The accretion rate then reaches its maximum value of $10^{-3} \ \mdot$ where it plateaus for 100 years (point 6). Here the star undergoes rapid expansion, with $\log_{10}{R/\text{R}_\odot}$ increasing from $0.4$ to roughly $2.0$, while simultaneously cooling down from $\log_{10}{T_\text{eff}/\text{K}}=4.6$ to $3.8$. During the plateau the star also maintains a nearly constant luminosity. As it enters the decay phase (point 7) the star begins to contract and heat up. The resulting trajectory is a curve towards the position it had before the burst started. Point 8 marks the termination of the \texttt{burst\_model} and the beginning of the \texttt{post\_burst\_model}. This model also captures the ramp-down of the accretion rate before the ZAMS mass is reached, and is terminated when the central hydrogen abundance is below 0.5 (point 9). 

\begin{figure}[h!]
    \centering
    \includegraphics[width=1.0\linewidth]{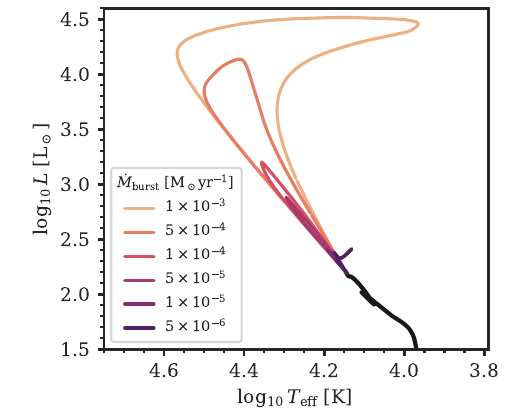}
    \caption{Evolutionary tracks for all calculated models undergoing a single accretion burst. The burst magnitudes span from $5\times10^{-6}$ to $10^{-3} \ \mdot$. Compared to Fig.~\ref{fig:HR-diagram}, this figure is restricted to the burst phase.}
    \label{fig:HR_diagram_all_tracks}
\end{figure}

We also calculated models with burst magnitudes 
$\dot{M}_\text{burst}=5\times10^{-6}, \ 10^{-5}, \ 5\times10^{-5}, \ 10^{-4}, \ \text{and} \ 5\times10^{-4} \ \mdot$. Fig. ~\ref{fig:HR_diagram_all_tracks} shows the corresponding evolutionary tracks, zoomed in on the burst phase. The pre-burst evolution is shown in black, while tracks of increasing burst magnitude are colored from dark purple to yellow.

We note that all models exhibit qualitatively similar burst excursions in the HR-diagram. Lower-magnitude bursts produce more modest deviations from the pre-burst track, while higher-magnitude bursts show pronounced luminosity increases followed by envelope expansion at nearly constant luminosity, leading to a decrease in effective temperature. As the accretion rate decays, the tracks return toward their pre-burst positions. Additionally, the excursion tracks of our intermediate-mass pre-MS stars are qualitatively similar to those  of low-mass protostars reported by \cite{Elbakyan2019} using the \texttt{STELLAR} code \citep{Yorke_and_Bodenheimer2008}, but are distinct from the excursion tracks of massive young stellar objects found in \cite{Meyer2019}---where variable accretion rates where adopted---using the \texttt{GENEVA} code \citep{Eggenberger2008}.

\subsection{Internal structures}\label{subsec:internal_structures}

Our analysis of the internal structures of our models is two-fold. First, we analyze the receding convective cores of our models during the bursts. This behavior is observed across all the models, but is more dramatic for higher burst magnitudes. Second, we characterize the chemical discontinuity left behind by the receding convective core.

\subsubsection{Receding convective cores}

\begin{figure}[h!]
    \centering
    \includegraphics[width=1.0\linewidth]{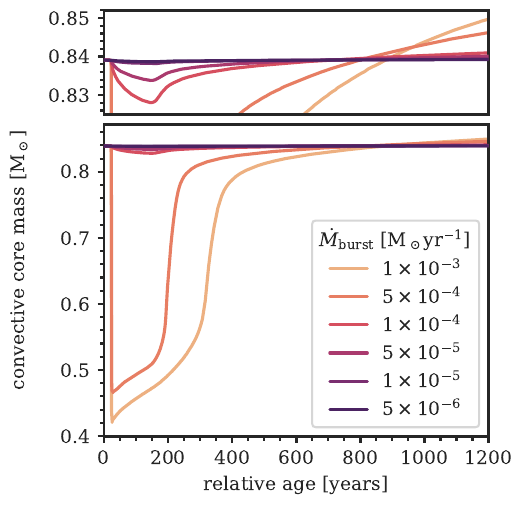}
    \caption{Evolution of convective core masses for all burst magnitudes considered. Relative age is calculated relative to the beginning of the \texttt{burst\_model}. The retraction of the core is substantially more dramatic for the two greatest burst magnitudes ($10^{-3}$ and $5\times10^{-4} \ \mdot$), hence, the top panel is zoomed in on the lesser burst magnitudes.}
    \label{fig:convective_core_mass_all_bursts}
\end{figure}

At the burst onset, the deep interiors of the models initially remain mostly unchanged, however, as the accretion rate plateaus at the burst maximum, we observe receding convective cores in all our models. This behavior is illustrated in Fig. \ref{fig:convective_core_mass_all_bursts} where the evolutions of the convective core masses are drawn. Each line drawn corresponds to a different \texttt{burst\_model} with a unique burst magnitude. Time is calculated from the initial age of the \texttt{burst\_model}, here denoted relative age (henceforth, relative age will always refer to time since beginning of \texttt{burst\_model}). As mentioned, the effects showcased here are more dramatic for higher burst magnitudes (bottom panel). The top panel is therefore zoomed in on the lesser burst magnitudes.

The convective core of the model undergoing the greatest burst ($10^{-3} \ \mdot$) is roughly halved in mass, receding from $0.84 \ \msun$ to $0.41 \ \msun$. In contrast, the model undergoing a $10^{-4} \ \mdot$ burst shows a much smaller reduction of only $0.01 \ \msun$. The maximum retreats of the convective cores across all our models are summarized in Table \ref{tab:burst}. 

\begin{figure*}[h!]
    \centering
    \includegraphics[width=1.0\linewidth]{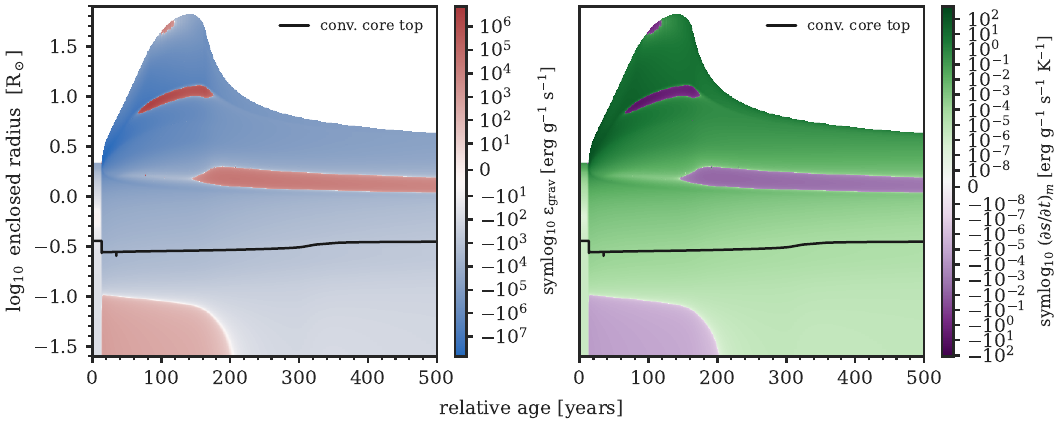}
    \caption{2D maps of $\egrav$ (left panel) and rate of specific entropy (right panel) across the burst. The maps are encompassed by enclosed radius and time axes. In the left panel blue colors correspond to $\egrav<0$ and red colors $\egrav>0$. In the right panel purple colors correspond to $(\partial s/\partial t)_m<0$ and green colors $(\partial s/\partial t)_m>0$.}
    \label{fig:eps_grav_heatmap}
\end{figure*}

From examining $\egrav$ (Sect. \ref{sec:accretion_in_mesa}), we can see what causes the core to recede during the burst. This is illustrated in Fig. \ref{fig:eps_grav_heatmap} where the surface radius of the model is drawn against relative age, and with the interior colored by $\egrav$ (left) and $(\partial s/\partial t)_m$ (right). $\egrav$ is calculated from Eq. (\ref{eq:egrav_paxton_2018}) and displayed on a symmetric base-10 logarithmic scale, taking linear values between $-10$ and $10$. The rate of specific entropy is calculated as, $(\partial s/\partial t)_m=-\egrav/T$, and similarly displayed on a symmetric scale taking linear values between $-10^{-8}$ and $10^{-8}$. The extent of the convective core is also drawn as a black line.

We can examine both $\egrav$ and $(\partial s/\partial t)_m$ to see which regions expand and which contract (Sect. \ref{sec:accretion_in_mesa}). Starting with $\egrav$, this quantity appears due to an imbalance between the luminosity gradient ($\partial L/\partial m$) and the net nuclear energy generation rate ($\epsilon_{\text{nuc}}$; Eq. \ref{eq:energy_equation_with_eps_grav_and_dsdt}). When $\egrav$ is positive a layer will contract and release gravothermal heat, and when it is negative a layer expands and gains gravothermal energy. This is reflected in the left panel of Fig. \ref{fig:eps_grav_heatmap}, where because of the time-scale hierarchy during the burst (Sec. \ref{sec:accretion_in_mesa}; $\tau_{\dot{M}}$ versus $\tau_{\text{KH}}$), the surface layers cannot accommodate the added energy. We therefore observe that they gain gravothermal energy and expand in the form negative $\egrav$. Conversely, between ages 0 and 200 years, we observe positive $\egrav$ in the core regions of the star. The accretion burst and the expansion of the envelope drive an energy imbalance at the center, producing positive $\egrav$ values as the core contracts. 

We also observe positive $\egrav$ from 200 years and onward at $\log_{10}{r/R_\odot}\simeq0.3$, and around $\log_{10}{r/R_\odot}\simeq1.0$ between ages 50 and 200 years. We attribute the former to steady off-center deuterium burning, which is evident when examining the deuterium abundance profile: from the surface towards the center the profile is constant at its primordial value, until $\log_{10}{r/R_\odot}\simeq0.3$ where it drops to zero. Simultaneously, there is an increase in $\hethree$ at this location (see Fig. \ref{app:fig:deuterium}). The pocket of positive $\egrav$ at $\log_{10}{r/R_\odot}\simeq1.0$ does not coincide with off-center burning, rather, we find that $\egrav$ takes on positive values because both the differential work and the internal energy rate decrease at $\log_{10}{r/R_\odot}\simeq1.0$ (see Fig. \ref{app:fig:components}). Additionally, $\epsilon_{\dot{M}}$ (Sect. \ref{sec:MESA}) also takes on high values ($\approx5\times10^{7} \ \text{erg}\ \text{g}^{-1} \ \text{s}^{-1}$) at this location. This represents the case of long thermal timescales $(\tau_{\dot{M}}\ll\tau_\text{KH})$, where material is advected adiabatically through the stellar envelope before finally releasing its retained heat (\citetalias{Paxton2019}).

We arrive at similar conclusions when examining $(\partial s/\partial t)_m$ in the right panel: the outer layers become highly entropic and expand to accommodate the added entropy. In the center, the star loses entropy and contracts.

\subsubsection{Chemical stratification}\label{subsubsec:chemical_stratification}
\begin{figure}
    \centering
    \includegraphics[width=1.0\linewidth]{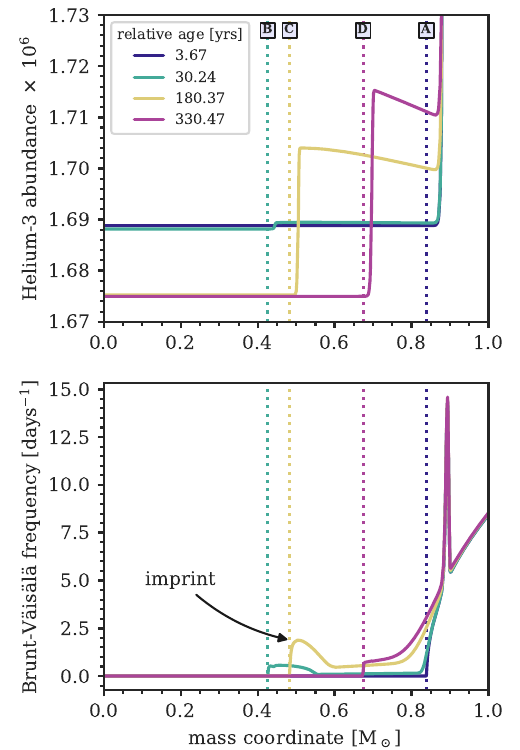}
    \caption{Top panel shows $\hethree$ abundance profiles for four selected structure profiles of the model undergoing $\dot{M}_\text{burst}=10^{-3} \ \mdot$. Model A (dark blue) at 3.67 years is taken just before the burst is initiated. Model B (green) is during burst, and models C (yellow) and D (purple) represent later stages as the core is re-establishing itself. Bottom panel shows a zoom-in on the Brunt-Väisälä profiles. The self-named asteroseismic imprint from the accretion burst is indicated by an arrow. Dotted lines indicate the extent of the convective core for each model.}
    \label{fig:chemistry}
\end{figure}

We illustrate the chemical trace left behind by the receding convective core in Fig. \ref{fig:chemistry}. From the model undergoing the greatest burst magnitude ($10^{-3} \ \mdot$) we selected four structure profiles (models A, B, C, and D) across the \texttt{burst\_model}. The ages of the models are calculated relative to the beginning of the \texttt{burst\_model}. Across all three panels in Fig. \ref{fig:chemistry}, the models are identified as follows: Model A (3.57 years; dark blue) represents the structure moments before the burst is initiated. Model B (30.24 years; green) shows the structure at the point where the convective core has receded to its maximum depth. Model C (180.37 years; yellow) captures a stage at which the core is mixing back towards its original position, erasing the chemical stratification. Finally, model D (330.47 years; purple) illustrates a later stage of the re-mixing as the initial size of the convective core is nearly restored. The top panel shows a zoom-in on the $\hethree$ abundance profiles for all four models. Dotted lines in all panels show the extent of the convective core for each stage. For the following, we will focus only on this top panel, and discuss the bottom panel in Sect. \ref{subsec:asteroseismic_analysis}.

Initially, at model A (dark blue), the $\hethree$ model (full line) is flat throughout the convective core (dotted line). At model B (green) the convective core has receded, and a small discontinuity in the abundance profile of around $\Delta\hethree\approx0.002\times10^{-6}$ is observed just outside the core-boundary. In model C (yellow) the discontinuity grows, and by model D the difference between the core and the core-boundary in the discontinuity is $\Delta\hethree\approx0.04\times10^{-6}$. We note that several  elements exhibit this behavior, but among these $\hethree$ was the most abundant, and therefore used as a tracer for chemical stratification and shown in Fig. \ref{fig:chemistry}. Other elements that contribute to the chemical stratification are $\liseven$, $\beseven$, and $\nififteen$.  

\subsection{Asteroseismic imprint}\label{subsec:asteroseismic_analysis}

In this section we asses the asteroseismic imprints of the accretion burst. In Sect. \ref{subsec:bv} we showcase Brunt-Väisälä frequency profiles altered by the receding convective core (a discussion on the changes in the outer layers is reserved for Appendix \ref{app:fig:brunt_N_evolution}). In Sect. \ref{sec:PSP} we first draw fundamental period-spacings ($\Pi_0$) for all our models, and show how these evolve across bursts of different magnitudes. We then employ \texttt{GYRE} to calculate individual \textit{g}-mode pulsation frequencies and compare period-spacing diagrams between burst models and constant accretion models. In Sect. \ref{subsec:FT_of_PSP} we characterize the role the imprint from the burst plays in the observed patterns by using the Fourier transform of the period-spacing pattern, and furthermore examine eigenfunctions of selected pulsation modes.

\subsubsection{Brunt-Väisälä frequency profiles}\label{subsec:bv}

The Brunt-Väisälä frequency (\citealp[\textit{N};][]{Vaisala1925}) is a useful quantity when examining \textit{g}-mode pulsations. In the simplest interpretation, \textit{N} is the frequency of oscillation of a perturbed parcel of fluid in a convectively stable environment \citep{HansenBook2004}. Oscillatory motion is attained when gravity, or buoyancy, attempts to restore the perturbed parcel to its equilibrium position. This is in essence \textit{g}-modes. In terms of the Brunt-Väisälä frequency, this criterion reduces to $N^2>0$. When this criterion is not met ($N^2<0$), a perturbed parcel finds itself in a convectively unstable environment and will rise or sink until it breaks up. The squared Brunt-Väisälä frequency, in the case of an idealized, fully ionized gas, reads \citep{Miglio2008},
\begin{equation}\label{eq:brunt_N2}
    N^2 \simeq \frac{g^2\rho}{p}\left(\nabla_\text{ad}-\nabla_T+\nabla_\mu\right)
\end{equation}
where \textit{g}, $\rho$, and \textit{p} are local values of gravity, density and pressure. Furthermore, $\nabla_T$ is the actual temperature gradient, $\nabla_\text{ad}$ is the adiabatic temperature gradient, and $\nabla_\mu$ is the chemical gradient. These are defined as,
\begin{equation}
    \nabla_T = \frac{\text{d} \ln{T}}{\text{d}\ln{p}},\quad \nabla_\text{ad}=\left(\frac{\partial \ln{T}}{\partial\ln{p}}\right)_S, \quad \nabla_\mu=\frac{\text{d} \ln{\mu}}{\text{d}\ln{p}},
\end{equation}
where $T$ and $\mu$ are local values of temperature and the mean molecular weight, and the subscript \textit{S} indicates constant entropy.
In principle, the parenthesis with the gradients state the Ledoux criteria for convection. The run of $N^2$ against, for example, mass coordinates will therefore tell which regions in the star are convectively unstable ($N^2<0$), through which no \textit{g}-modes propagate, and which regions are convectively stable ($N^2>0$), where \textit{g}-modes can propagate. Such a diagram is commonly called a propagation diagram \citep{AertsBook2010} as it illustrates regions where modes of oscillation exist. 

In the bottom panel of Fig. \ref{fig:chemistry} the Brunt-Väisälä frequency ($N$) is drawn against mass coordinate. Notably, negative $N^2$-values have been truncated to zero. The ages, naming, and coloring of the profiles are the same as in the top panel (see Sect. \ref{subsubsec:chemical_stratification}). The boundary of the convective core in each model is similarly indicated by dotted lines.

% \begin{figure}[h!]
%     \centering
%     \includegraphics[width=1.0\linewidth]{updated_figures/brunt_N2_with_burning.pdf}
%     \caption{Brunt-Väisälä frequency profiles (black) and burning profiles of both the \textit{pp}-chain (orange) and the CNO-cycle (dot-dot-dashed; red) for the model undergoing the greatest burst magnitude ($10^{-3} \ \mdot$). Taken at a relative age of 3.57 years, the top panel illustrates the profiles just before the onset of the burst, that is, model A in Fig. \ref{fig:chemistry}. The bottom panel corresponds to model C in Fig. \ref{fig:chemistry} where the convective core is re-mixing back to its original position. A steep drop in the \textit{pp}-burning profile is highlighted, and also a bump in the Brunt-Väisälä profile, here denoted imprint. Both of these features live just outside the convective core, the boundary of which is drawn as a gray dashed line.}
%     \label{fig:brunt_N2_with_burning}
% \end{figure}

For model A (dark blue), $N$ is zero up to approximately $0.83 \ \msun$, which is exactly the extent of the convective core at this stage. Beyond this point $N$ takes positive non-zero values. This corresponds to the radiative layers of the star, and here \textit{g}-modes can propagate. At the coordinate of roughly $0.9 \ \msun$, $N$ takes large values in the form of a spike (in fact, the spike is present in at all four stages A, B, C, and D). Comparing with the top panel, we observe that the spike coincides with a sharp discontinuity in the $\hethree$ profile. As is evident from Eq. (\ref{eq:brunt_N2}), such discontinuities will affect $N$. For model B (green), we observe that $N$ takes zero-values only up to approximately $0.43 \ \msun$. This is a result of the recently receded convective core. Beyond the core the small discontinuity in $\hethree$ is present, and consequently $N$ rises to about $0.5 \ \text{days}^{-1}$ before decreasing again at a coordinate of $0.55 \ \msun$. We labeled this as the "imprint" in Fig. \ref{fig:chemistry} (see arrow with text). The imprint grows to roughly $2.0 \ \text{days}^{-1}$ (model C; yellow) before fading away as the core extends again in mass coordinate (model D; purple).  
% The different shapes of $N$ will give different pulsation patterns.

\subsubsection{Period-spacing diagrams}\label{sec:PSP}
Computation of individual eigenmodes are carried out with the stellar oscillation code \texttt{GYRE}. From here we obtain the periods of the oscillations, allowing us to draw so-called period-spacing diagrams. Here the distance between periods of subsequent mode order are drawn against the actual periods. Asymptotic analysis shows that low-degree, high-order \textit{g}-modes are approximately equally spaced with a period-spacing given by \citep{Tassoul1980, AertsBook2010},
\begin{equation}\label{eq:Pi0}
    \Delta \Pi_{\ell}=\frac{2\pi^2}{\sqrt{\ell(\ell+1)}}\left(\int_{r_1}^{r_2}N\frac{dr}{r}\right)^{-1},
\end{equation}
where the integral represents radial integration over the entire \textit{g}-mode cavity, and $\ell$ is the spherical degree. The equal spacing is broken by the presence of sharp chemical discontinuities which create consecutive dips in the period-spacing in the absence of rotation, magnetic fields, and mode interactions \citep{Miglio2008}.

\begin{figure}[h!]
    \centering
    \includegraphics[width=1.0\linewidth]{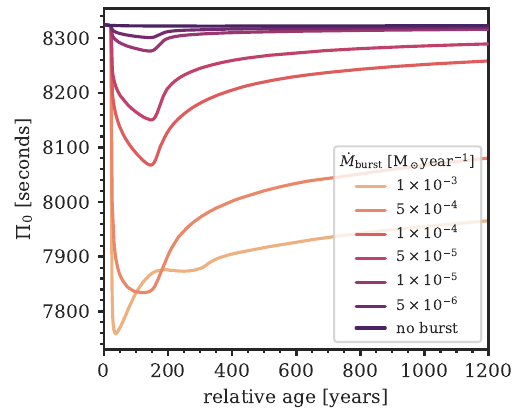}
    \caption{Evolution of the fundamental period spacing ($\Pi_0$) for all burst magnitudes considered. Relative age is calculated relative to the beginning of the \texttt{burst\_model}.}
    \label{fig:Pi0}
\end{figure}

Fig. \ref{fig:Pi0} shows the evolution of $\Pi_0 = 2\pi^2\left(\int_{r_1}^{r_2}N/r \ dr\right)^{-1}$, in other words, the degree-independent, fundamental spacing, or, buoyancy travel time, calculated with Eq. (\ref{eq:Pi0}). Models undergoing increasingly greater burst magnitudes are distinguished by color, increasing from dark purple to yellow.

We observe across all models $\Pi_0$ to decrease during the burst, and gradually increase once the burst is over. The effect is again most significant for the model undergoing the burst of $10^{-3} \ \mdot$. Here $\Pi_0$ drops from roughly 8420 to 7760 seconds. The $\Pi_0$ values in Fig. \ref{fig:Pi0} are generally lower than values expected from zero-burst evolution, and may therefore provide evidence as to whether a pre-MS \textit{g}-mode pulsator has undergone one or more bursts during its formation. Similar conclusions are drawn in \cite{Rui2021}, specifically for merger remnants in their red-giant stage of evolution. The origins of the changes $\Pi_0$ are the changes in the shape of the Brunt-Väisälä frequency profile during the burst, and also the expansion of the star. The latter enlarges the \textit{g}-mode cavity and influences $\Pi_0$ through the integral over $N/r$ (see Eq. \ref{eq:Pi0}). This is evident from Fig. \ref{app:fig:brunt_N_evolution} where we showcase the full shapes of the Brunt-Väisälä profiles for models A, B, C and D in Fig. \ref{fig:chemistry}

\begin{figure}[h!]
    \centering
    \includegraphics[width=1.0\linewidth]{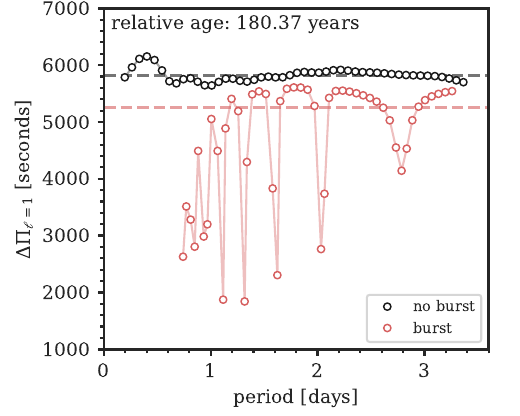}
    \caption{Period-spacing patterns of the model undergoing a burst of $10^{-3} \ \mdot$ (red), and a model evolving with a constant rate (black). The models are captured at the same age of 180.37 years after the beginning of the \texttt{burst\_model}. The red model is thus model C in Fig. \ref{fig:chemistry}.}
    \label{fig:period_spacing_result}
\end{figure}

We show in Fig. \ref{fig:period_spacing_result} period-spacing diagrams of two models; one undergoing no accretion burst (black; Sect. \ref{subsec:constant_rate}) and another undergoing an accretion burst of $10^{-3} \ \mdot$ (red). Both models are captured at an age corresponding to roughly 180 years after the beginning of the \texttt{burst\_model} (model C in Fig. \ref{fig:chemistry}).

The pattern of the no-burst model reflects what we would expect of a non-rotating, non-magnetic star with a smooth Brunt-Väisälä profile in the asymptotic limit. It is a nearly horizontal pattern indicating the modes are more or less equally spaced by a fundamental spacing (black dashed line). Examining the burst model, however, we observe a different pattern. Outside occasional even spacings (indicated by a red dashed line) we observe quasi-periodic dips in the period-spacing, reminiscent of patterns seen with modes trapped by chemical discontinuities in the stellar interior \citep{Miglio2008}. These dips are a consequence of the receding convective core during the burst, yet some of them may originate from outer regions in the star \citep{Montgomery2003}. We elaborate upon this in Sect. \ref{subsec:FT_of_PSP}.

We summarize in Table \ref{tab:burst} the following quantities: We visually inspect the period-spacing patterns of all our models ($\dot{M}_\text{burst}$; first column) and note the approximate times ($\tau_{\mathrm{imprint}}$; third column) when the signatures from the bursts disappear. For $\dot{M}_\text{burst}<5\times10^{-5} \ \mdot$ no discernible signature is seen, hence these entries are left empty. We also counted the maximum number of dips occurring in the period-spacing patterns during the bursts ($N_\text{dips}$; fourth column), and the maximum retreat of the convective cores ($\Delta M_{\mathrm{cc,max}}$; second column). For $\dot{M}_\text{burst} = 10^{-4}$ and $5\times10^{-5} \ \mdot$, while a single dip is counted in each, these are more shallow than those for $\dot{M}_\text{burst}\geq5\times10^{-4} \ \mdot$.

\begin{table}[htbp]
\centering
\caption{Maximum retreat of the convective cores ($\Delta M_{\mathrm{cc}}$), lifetime of the imprints ($\tau_{\mathrm{imprint}}$; approximate value noted by eye), and number of dips in the period-spacing patterns ($N_\text{dips}$) across all our models (identified by $\dot{M}_\text{burst}$).}
\label{tab:burst}
\begin{tabular}{lccc}
\toprule
$\dot{M}_\text{burst} \ [\mdot]$ & $\Delta M_{\mathrm{cc}} \ [\msun]$ & $\tau_{\mathrm{imprint}} \ [\text{yrs}]$ & $N_\text{dips}$\\
\midrule
$1\times10^{-3}$ & 0.42 & 475 & 10 \\
$5\times10^{-4}$ & 0.37 & 250 & 2 \\
$1\times10^{-4}$ & 0.01 & 200 & 1 \\
$5\times10^{-5}$ & $5.4\times10^{-3}$ & 150 & 1 \\
$1\times10^{-5}$ & $9.6\times10^{-4}$ &  &  \\
$5\times10^{-6}$ & $4.6\times10^{-4}$ &  &  \\
\bottomrule
\end{tabular}
\end{table}

\subsubsection{Assessing the origins of the signatures in the period-spacing patterns}\label{subsec:FT_of_PSP}
\begin{figure}[h!]
    \centering
    \includegraphics[width=1.0\linewidth]{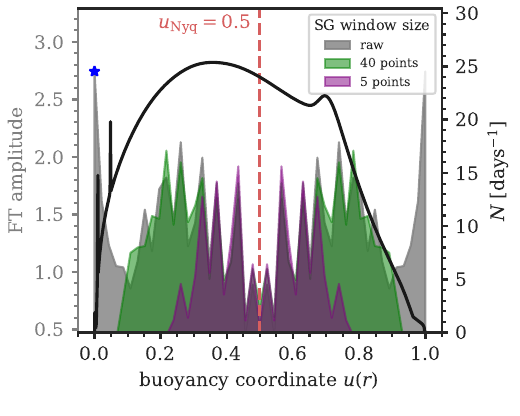}
    \caption{Brunt-Väisälä frequency ($N$; black) and Fourier transform of the period-spacing pattern (FT amplitude). The different colors of the FT (gray, green, purple) indicate how much the $\Delta \Pi^{\text{norm}}_{\ell=1}(n)$ has been smoothed before calculating the FT. The Nyquist limit of the FT is indicated by a dashed red line. A blue star indicates the most prominent peak below $u(r)<0.5$ before any smoothing is applied. The model selected for this figure is model C of Fig. \ref{fig:chemistry}.}
    \label{fig:FFT_PSP}
\end{figure}

Recent advances in the interpretation of \textit{g}-mode period-spacings make use of the Fourier transform (FT) of the period-spacing pattern itself \citep{Guo2025, Guo2026, Wu2026, Henneco2026}. Hence, we calculate the buoyancy coordinate $u(r)$ as,
\begin{equation}\label{eq:u_r}
    u(r)=\frac{\int_{r_1}^{r}N(r')/r' \ \text{d}r'}{\int_{r_1}^{r_2}N(r')/r' \ \text{d}r'} = \frac{\Pi^{-1}_r}{\Pi_0^{-1}},
\end{equation}
where $N(r)$ is the Brunt-Väisälä profile, $r$ is the radial coordinate, $r_1$ and $r_2$ and are the inner and outer boundaries of the \textit{g}-mode cavity. Segments with higher $N$-values at lower radial coordinates will contribute more to the integral in Eq. (\ref{eq:u_r}), and the effect of this is to stretch the Brunt-Väisälä profile in the buoyancy coordinate $u(r)$ and emphasize features from, for example, chemical discontinuities. Next we calculate the FT of the normalized period-spacing pattern, given as,
\begin{equation}\label{eq:FT_PSP}
    \Delta \Pi^{\text{norm}}_{\ell=1}(n) = \frac{\Delta \Pi_{\ell=1}(n)-\Delta\Pi_{\ell=1}}{\Delta\Pi_{\ell=1}},
\end{equation}
where $\Delta \Pi_{\ell=1}(n)$ are the dipole period-spacings as a function of radial order $n$, and $\Delta\Pi_{\ell=1}$ is the fundamental spacing of dipole modes, calculated as,
\begin{equation}\label{eq:Pi1}
    \Delta\Pi_{\ell=1} = \frac{\Pi_0}{\sqrt{\ell(\ell+1)}} = \frac{\Pi_0}{\sqrt{2}}.
\end{equation}
In Fig. \ref{fig:FFT_PSP} we draw the Brunt-Väisälä profile (black) and the FT of the period-spacing pattern (gray) against the buoyancy coordinate, $u(r)$. Following \citet{Montgomery2003}, we use the relation $\Delta n = 1/u(r)$ to first associate a spectral coordinate in $u(r)$ with specific features in the Brunt-Väisälä frequency profile. These features then serve as the physical source of the observed variability in the period-spacing pattern across multiple mode orders $\Delta n$. The Nyquist limit of the FT is $u(r)=1/(2\langle n_{i}-n_{i-1}\rangle)=0.5$ (since $n_{i}-n_{i-1}=1.0$ is the spacing between subsequent mode orders) and the FT is mirrored around this value ($u_\text{Nyq}$; red dashed line). However, FT-peaks in the region $u(r)<0.5$ may be aliases from real physical features at $u(r)>0.5$, hence, we also draw the super-Nyquist regime \citep{Montgomery2003,Henneco2026,Wu2026}. The model selected for Fig. \ref{fig:FFT_PSP} is model C of Fig. \ref{fig:chemistry}. 
% In doing so, we apply $\Delta n=1/u(r)$, where $\Delta n$ refers to how many mode orders the variability in the period-spacing pattern covers \citep{Montgomery2003}. We are thus able to identify significant peaks in the FT of the period-spacing pattern with features in the Brunt-Väisälä profile.

The imprint of the core retraction in Fig. \ref{fig:chemistry} does not contribute significantly to the integral in Eq. (\ref{eq:u_r}), hence this feature is not stretched when drawing $N$ against $u(r)$ in Fig. \ref{fig:FFT_PSP}. Examining the FT of the period-spacing pattern, we observe one prominent peak (blue star) at essentially $u(r)=0.0$ (mirrored at $u(r)=1.0$). This location corresponds to the convective core boundary, and is where the imprint from the burst lies. We may therefore argue that parts of the variability in the period-spacing pattern originate from the imprint of the accretion burst (see bottom panel of Fig. \ref{fig:chemistry}). The location of the peak also indicates that the variability covers close to 50 mode orders (since we calculate 50 modes, $u(r)\approx0.0 \Leftrightarrow \Delta n\approx50$). Therefore the imprint near the core should manifest as a global trend in the period-spacing pattern.

We now smooth out long-term trends in the period-spacing pattern using a Savitzky-Golay filter \citep{SG_filter}; first with a window size of 40 data points (green), and then 5 (purple). In the first step, the peak at $u(r)=0.0$ disappears. Visual inspection of the smoothed period-spacing pattern (Fig. \ref{fig:app:smoothed_psp}) reveals that the smoothing effectively removes long-term trends; what remains are the dips crossing the zero-point of $\Delta P^{\text{norm}}_{\ell=1}(n)$ periodically. In the next step, the smoothing attenuates the dips seen at higher periods, leaving the variability at low periods as the dominant. The resulting FTs in Fig. \ref{fig:FFT_PSP} share the same peaks, but are truncated at increasingly higher $u(r)$ as the period-spacing pattern is smoothed. 

Evidently, the imprint at the core-boundary cannot alone explain the period-spacing pattern of Fig. \ref{fig:period_spacing_result}. Instead, we may suspect that additional components originate from the features seen in Fig. \ref{fig:app:BV_surf}, in the outer regions of the star. The peaks in the FT representing short-term variability show up in the range $0.1\leq u(r) \leq0.5$, but may also originate from real alias peaks in the super-Nyquist regime ($u(r)>u_\text{Nyq}$; \citealp[]{Montgomery2003, Miglio2008, Wu2026}). The Brunt-Väisälä features discussed are shown in Fig. \ref{fig:app:smoothed_psp}; we may speculate that they produce real FT-components reflected back into the sub-Nyquist regime \citep{Henneco2026}.

\begin{figure}
    \centering
    \includegraphics[width=1.0\linewidth]{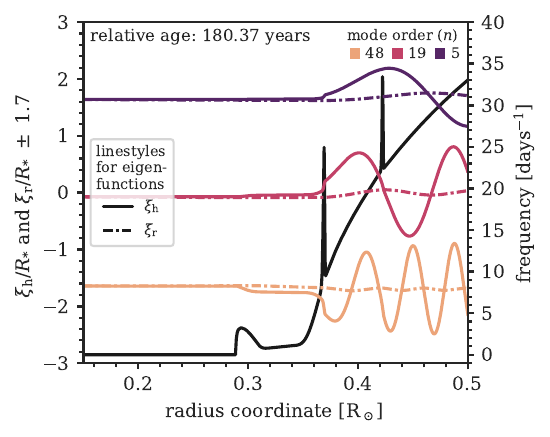}
    \caption{Eigenfunctions of stellar oscillations (left y-axis) and Brunt-Väisälä frequency (right y-axis) drawn against radius coordinates in the range $0.15 <r \ [\rsun]<0.5$. The Brunt-Väisälä profile is drawn in black, while the horizontal- and radial displacements ($\xi_\text{h}$ and $\xi_\text{r}$) of the eigenfunctions of selected modes ($n=5,19,48$) are drawn as full- and dash-dotted lines respectively (see leftmost legend). And offset of $\pm1.7$ has been applied to $n=5$ and $n=48$ in order to distinguish all three modes.}
    \label{fig:prop_with_eigenmodes}
\end{figure}

Additional information is obtained from comparing the widths of peaks (glitches, spikes and imprints) in \textit{N} ($\delta N_\text{peak}$) with the local wavelengths of the eigenmodes of the stellar oscillations \citep[$\lambda_\text{local}$;][]{Henneco2025}. If $\lambda_\text{local}>\delta N_\text{peak}$ at the location of our imprint we can expect the peak to perturb the eigenmodes \citep{Cunha2020}. This is illustrated in Fig. \ref{fig:prop_with_eigenmodes} where eigenfunctions (left y-axis) and frequency components (right y-axis) are drawn against radius coordinates zoomed in on the convective-core-boundary region. $N$ is drawn in black and eigenfunctions of modes of radial order $n=5,19$, and $48$ are indicated by colors. Specifically, the horizontal- and radial displacements ($\xi_\text{h}$ and $\xi_\text{r}$) are drawn as full- and dash-dotted lines respectively. The modes have the following frequency values ($\nu_n$): $\nu_5=0.504 \ \text{d}^{-1}$, $\nu_{19}=0.899 \ \text{d}^{-1}$, and $\nu_{48}=1.895 \ \text{d}^{-1}$. The $n=5$ and $n=48$ eigenfunctions have been offset by $\pm1.7$ in order to distinguish all three modes.

In Fig. \ref{fig:prop_with_eigenmodes} we find three peaks in \textit{N}; the imprint located at $0.3 \ \rsun$, and the chemical peaks at $0.37 \ \rsun$ and $0.42 \ \rsun$. With respective widths of approximately $0.1 \ \rsun$, $0.01 \ \rsun$, and $0.01 \ \rsun$ we are in the regime where each peak should perturb the waves (since $\lambda_\text{local}>\delta N_\text{peak}$). This is also observed; one can note small perturbations to the waves as they traverse the Brunt-Väisälä peaks. The conclusions from the discussion above, however, still stand. The variability in the period-spacing pattern from the imprint will manifest as a global trend.

\section{Discussion}\label{sec:discussion}
In this section we assess the validity of our burst model, examine the impact of key modeling assumptions, and outline directions for future work.

\subsection{Global burst properties}
In Sect. \ref{subsec:accretion_burst} we argued that our toy model of the bursts was representative of actual FUor-like bursts. We may validate our assumption by considering the resulting stellar evolution using this model. Firstly, we note that the duration of the plateau ($t_\text{plateau}$) phase is broadly consistent with that of FU~Orionis and V883~Ori (\citealp[see Table~A.1 in][]{VorobyovFUORS-2021}) and the peak accretion rate is consistent with the MRI-triggered bursts for intermediate-mass protostars considered in this work \citep{Das2025}. The resulting trajectories in the HR-diagram (Figs. \ref{fig:HR-diagram} and \ref{fig:HR_diagram_all_tracks}) are consistent with the brightness variations observed in FUor-like bursts \citep{Hartmann1996, Audard2014, Semkov2018, Szab2021}. Specifically, the luminosity increases span from modest enhancements of a factor $\sim$$1.5$ for the weakest burst, up to $\sim$$250$ for the strongest burst. The variations for the strongest bursts are in line with FUor-type outbursts which show variability on the order of 3.5 to 6.0 magnitudes, with stronger bursts amounting to 8 magnitudes \citep{Lucas2020}. The weakest bursts are representative of the class of so-called accretion bursters, showing variability of around 0.5 to 2.5 magnitudes \citep{Guo2020}.

In this work we focused only on burst shapes characterized by $t_\text{rise}=4 \ \text{yr}$, $t_\text{plateau}=100 \ \text{yr}$, and $t_\text{decay}=80 \ \text{yr}$ (Eq. \ref{eq:burst_profile_sigmoids}). Our \texttt{MESA} setup allows for more burst profiles to be explored, which will be the subject of future work. Furthermore, our models undergo only one accretion burst. More sophisticated simulations and observations of knots in jets suggest that protostars experience multiple occasional bursts \citep{Bae2014, VorobyovBasu2015, Vorobyov2018, Cecil2024, Nayakshin2024, Hartmann2025}. In such cases, we can speculate that the resulting asteroseismic signatures may be both stronger and longer-lived than predicted here.

\subsection{Stellar models outlook}
Our choice to begin the accretion phase from initial seeds of $0.01 \ \text{M}_\odot$ and $1.5 \ \text{R}_\odot$ is motivated mainly by numerical stability \citep{Kunitomo2017}, as problems can arise in \texttt{MESA}'s equation of state and opacities at the seed conditions. However, the protostellar seed mass may be somewhat smaller (\citealp[$4\times10^{-3} \ \msun$;][]{Ahmad2023}). The consequences of different seed conditions are discussed in, for example, \cite{Joerjensen2025}. To summarize, our results are constistent only in the framework of beginning the accretion phase from a seed of $0.01 \ \text{M}_\odot$, but may withstand variations in the seed radius.

The tendency of our models to expand is consistent with previous studies. Mass transfer onto intermediate-mass main-sequence stars often leads to rapid expansion \citep{Kippenhahn1977, Hamstars}, while fully convective models can contract even under moderate accretion rates \citep{Prialnik1985}. The response is mostly sensitive to the local entropy stratification. Radiative envelopes ($ds/dr>0$) tend to expand under mass accretion, whereas convective envelopes ($ds/dr=0$ or $<0$) tend to contract (\citealp[chapter 7.3 in][]{HansenBook2004}).

We emphasize that the accretion phase is highly idealized. Not only do the models undergo only one accretion burst, but they also evolve with a constant accretion rate. Accretion onto protostars is instead a variable and irregular process. This can be captured with \texttt{MESA} \citep{Steindl2022}, but was not the subject of this work.

Rotation was not included, although young stars are likely fast rotators \citep{Cody2017, Froebrich2021, Noroozi2024}. In one-dimensional stellar models, rotation introduces additional mixing processes that may alter the chemical signatures identified here \citep{Paxton2013}. It also affects the pulsation properties obtained from \texttt{GYRE}: rotational splitting lifts the degeneracy in azimuthal order $m$ \citep{AertsBook2010}, and the Coriolis force acting as an additional restoring force induces slopes in period-spacing patterns from which internal rotation can be inferred \citep{VanReeth2016, Zwintz2017}. While the inclusion of rotation is out of the scope of this work, we provide a small discussion on it in Appendix. \ref{app:rotation}.

This work assumes only cold accretion, in which the energy carried by accreted material is radiated away instantaneously at the surface. In contrast, hot accretion assumes that some of this energy may be absorbed by the protostar itself. It is not clear to what extent hot accretion plays a role during the accretion phase, especially during accretion bursts. For example, \cite{Baraffe2009} found that negligible amounts of accretion energy are required for models undergoing episodic accretion to reproduce the luminosity spread of young stellar objects (YSOs) in open clusters. On the other hand, \cite{Steindl2021} determined that accretion energy is required for constant-accretion models to reproduce surface gravities and effective temperatures of another set of YSOs. Furthermore, the manner in which the energy is distributed in models varies between authors \citep{Hosokawa2011, Jensen2018, Elbakyan2019, Steindl2022, Joerjensen2025}. The choice to ignore hot accretion in this study was therefore also a matter of simplicity, and we leave the inclusion of hot accretion into our analysis as the subject of future work.  

We encountered difficulties in resolving the receding convective core during the burst. In the end, this was solved by limiting the minimum cell-size under composition changes to $m/M = 10^{-4}$ (\texttt{min\_dq\_for\_xa}; Sect. \ref{subsec:stellar_models}). The extent to which our results depend on this choice is challenging to evaluate. With values $<10^{-4}$ we observed the convective core to fragment or even periodically grow and retract. It remains unclear whether these features are physical or numerical in origin; in either case, they could not be robustly reproduced. Potential improvements include more advanced treatments of convection \citep{Rees2024}, or using \texttt{TDC} \citep{Jermyn2023}. with even shorter timesteps. We note that $\texttt{min\_dq\_for\_xa}=10^{-4}$ is frequently adopted in \texttt{MESA}'s benchmark \texttt{test\_suite} models, particularly for high-mass stars ($20\leq\text{M} \ \left[\msun\right] \leq 200$).

\subsection{Asteroseismic analysis outlook}
$\gamma$-Doradus stars also form an important class of \textit{g}-mode pulsators on the pre-MS \citep{Zwintz2013:gamma_dor, Steindl2021, Zwintz2022}. These are stars of lower masses between $1.4$ and $2.0 \ \msun$, which show high-order \textit{g}-modes with periods from 0.3 to 3 days \citep{Kaye1999} excited by the convective flux-blocking mechanism \citep{Guzik2000, Dupret2005}. Stars in this mass-range are also predicted to have multiple accretion bursts throughout their formation \citep{VorobyovBasu2015}, and we may suspect similar imprints in their period-spacing patterns. This hypothesis remains to be tested with dedicated modeling.

The period-spacing patterns shown in Fig. \ref{fig:period_spacing_result} are obtained from purely adiabatic calculations of the stellar eigenfrequencies. \cite{Steindl2021} demonstrated, with the inclusion of non-adiabatic effects in constant-accretion models, that pre-MS SPB stars constitute their own class of \textit{g}-mode pulsators. In \texttt{GYRE}, non-adiabatic effects are modeled with the inclusion of the linearized equations for energy and radiative diffusion; constituing a sixth-order boundary value problem \citep{GYRE2018}, obtained by ignoring perturbations of convective heating- and cooling terms. Fully non-adiabatic calculations would in principle also allow us to determine whether a star can sustain \textit{g}-mode pulsations during an accretion burst. However, such an analysis is challenging due to several factors such as the treatment of convection and the increased computational cost \citep{GYRE2020}. Even though the inclusion of non-adiabatic effects can affect the eigenfrequencies, we confirm that no significant difference is found between the adiabatic and non-adiabatic \texttt{GYRE} computations of selected models. Our period-spacing patterns should therefore be robust against the inclusion of non-adiabatic effects.

\subsection{Asteroseismic imprint as an identifier for FUor outbursts}

As discussed in Sect. \ref{sec:introduction}, the population of known FUors is likely highly incomplete due to the observational challenge of detecting the outburst and identifying the source as a FUor. This incompleteness has significant consequences for our ability to determine important quantities like the duty cycle of outbursting YSOs and the number of outbursts a YSO is likely to undergo in its formation. For YSOs that are discovered mid-outburst, it can also be impossible to ascertain the peak accretion rate of the burst or the accretion state of the star pre-outburst.

For intermediate-mass sources, even determining that an object is a FUor by spectroscopic confirmation can be challenging. The highest-mass known FUor, Gaia 20bdk with an estimated mass of $2.5 \pm 0.5$~$\msun$, shows very little disk-like absorption in its visible wavelength spectrum, while its near-infrared spectrum matches that of a FUor precisely \citep{2025A&A...695A.130S, 2026arXiv260512795P}. The visible range spectrum of Gaia 20bdk appears to be dominated by wind absorption and emission, while the high central stellar mass produces a high maximum disk temperature near the star, which can also disrupt the ability to detect the typically expected disk absorption lines \citep{2009ApJ...694L..64Z,2023ApJ...958..140C}.

It may in fact be expected that the spectra of YIMS FUors would be wind-dominated, due to their high accretion rates. Historically, the mass outflow rate of a FUor in outburst, $\dot{M}_w$, has been estimated to be approximately $\dot{M}_w = 0.1 \ \mdot$ \citep{1993ApJ...402..623C}. In radiation magneto-hydrodynamical simulations of low mass FUor outbursts, the base of the wind is found to be optically thick close to the star, supporting the observation that the visible and ultraviolet spectra of the low mass FUors are wind-dominated \citep{2020MNRAS.495.3494Z}. At higher $\dot{M}$, the resulting greater $\dot{M}_w$ would result in large fractions of the inner disk being obscured by a highly optically thick wind, potentially even obscuring portions of the infrared spectrum.

In the case of Gaia 20bdk, the outburst was observed, so the fact it matches FU Ori well in the infrared solidifies its classification as a bonafide FUor. However, with only spectroscopic information, the source might be missed. The well-known FUor V1057 Cyg has been in outburst since 1970 \citep{Hartmann1996}. In 1996, its spectrum underwent significant evolution and has been wind-dominated at visible wavelengths, while its infrared molecular absorption has gotten steadily weaker \citep{Szab2021}. Within a decade, its spectrum could evade identification as FUor-like.

The distinct signature in the \textit{g}-mode pulsations presented in this work may circumvent the challenges in establishing the FUor-like nature of an object in outburst. Additionally, the signature itself places constraints on the characteristics of the burst itself, such as the peak-accretion rate.

\section{Summary and Conclusions}\label{sec:conclusion}

We have modeled $4.0 \ \msun$ stars grown from $0.01 \ \msun$ seeds via constant accretion, introducing a burst at $3.5 \ \msun$ to assess its impact on the asteroseismic properties of pre-MS SPB stars. The stellar evolution was computed using \texttt{MESA}, and asteroseismic properties were obtained with \texttt{GYRE}.

We conclude that a single accretion burst imprints a signature in a model's the period-spacing pattern, allowing it to be distinguished from models that have not undergone a burst. The signature arises from both the expansion of the envelope, but also structural changes in the deep interior, where the receding convective core leaves behind a chemical signature. As the core re-establishes itself, this signature is gradually erased and largely disappears within 475 years after the burst maximum. 

The distinct signature of a FUor outburst, like the altered \textit{g}-mode period-spacing patterns described above, may provide a means of identifying intermediate-mass FUors in outburst or shortly after their bursts. Confirming such signatures observationally would require continuous, long base-line monitoring of YSOs to resolve individual mode frequencies. Owing, in part, to the radius inflation expected during the outburst these objects have exceptional luminosities and a high star-to-disk luminosity ratio. This allows the central star of such a system to be observed during a burst more readily than in low-mass FUors. For these reasons, intermediate-mass FUors are excellent targets for high-precision, high-cadence photometry, and a confirmed detection may be realized with data from photometric space missions such as TESS (\citealp[The Transiting Exoplanet Survey;][]{Ricker2014}), or PLATO (\citealp[PLAnetary Transits and Oscillations of stars;][]{Rauer2025}). Additionally, they are easily identified in color-magnitude diagrams and can be selected to determine whether they possess the predicted period-spacing patterns. The dependence of the period-spacing pattern on the peak accretion rate during the burst also places a valuable constraint on the accretion rate itself. Moreover, these oscillations would give a direct indication of the impact of an outburst on the structure of the star, which today remains unknown due to the lack of known post-outburst FUors.

Finally, the task of asteroseismic modeling involves comparing computed eigenfrequencies of stellar models with observed frequencies to infer stellar parameters \citep{Aerts2021}. Although the signature identified here is short-lived, it demonstrates that asteroseismic modeling of pre-MS stars \citep{Zwintz2005, Zwintz2007, Guenther2007, Guenther2009, Steindl2022II} may depend on the ability to distinguish burst-induced features from pulsation features associated with quiescent phases \citep{ Steindl2022}. 

\begin{acknowledgements}
We are grateful for the referee’s careful assessment of the manuscript and their helpful suggestions for its improvement. We thank the developers of \texttt{MESA} and \texttt{GYRE} for providing and maintaining the publicly available stellar structure and stellar pulsation codes. Johannes Jørgensen is grateful to Dominic Bowman for help in interpreting the gravity-mode period-spacing patterns, and also gives thanks to Joey Mombarg for providing the \texttt{MESA} routine to calculate $\Pi_0$. He also gives thanks to Earl Bellinger, Yale University, for assistance in the early stages of this work. Jan Henneco gratefully acknowledges UK Research and Innovation (UKRI) in the form of a Frontier Research grant under the UK government's ERC Horizon Europe funding guarantee (SYMPHONY; PI Bowman; grant number: EP/Y031059/1). The computational results presented here have been achieved (in part) using the LEO HPC infrastructure of the University of Innsbruck. This research has made use of Matplotlib \citep{Matplotlib}; NumPy \citep{numpy}; Seaborn \citep{seaborn}; Astropy \citep{astropy}; \texttt{MESA} SDK for Linux \citep{MESASDK2024}.
\end{acknowledgements}


\bibliographystyle{aa}
\bibliography{references}

\begin{appendix}

\section{Numerical tests of burst models}\label{app:numerical_test_single_burst} 

In this section we present results from numerical convergence tests of our burst models. We showcase only models for which the burst magnitude is $\dot{M}_\text{burst}=10^{-3} \ \text{M}_\odot \ \text{yr}^{-1}$ as these are the most demanding and the ones for which numerical convergence is challenging to attain.

\subsection{Increasing resolution}\label{Appendix A:Increasing resolution}
\begin{figure}[h!]
    \centering
    \includegraphics[width=1.0\linewidth]{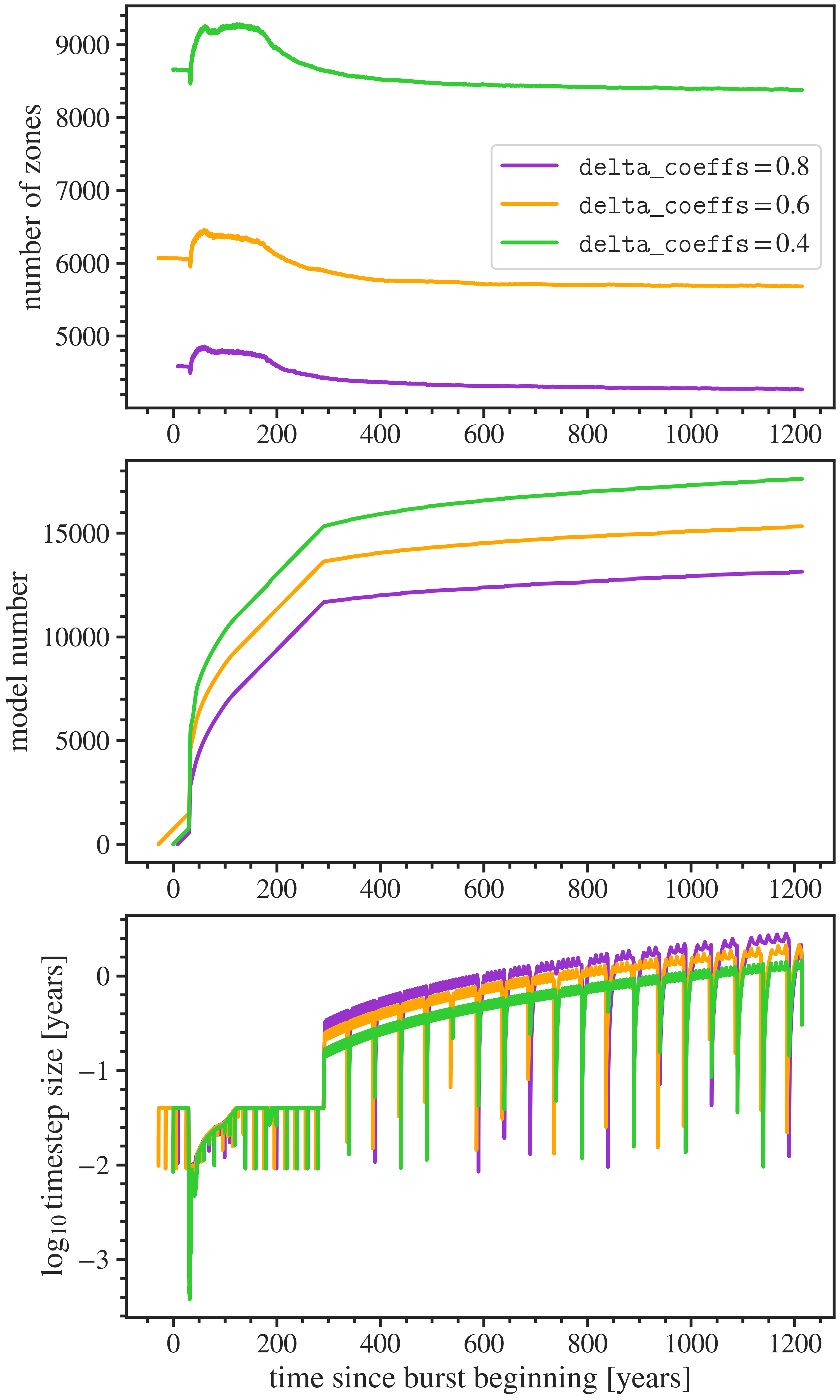}
    \caption{Model quantities affected by changes to the resolution settings in \texttt{MESA}. The coarsest, finer, and finest models respectively have $\texttt{*\_delta\_coeff}$'s of 0.8, 0.6, and 0.4, and are drawn in purple, orange and green. Top panel illustrates number of zones, middle panel illustrates model number, and bottom panel illustrates timestep size in years. All the quantities are drawn against time since burst beginning.}
    \label{fig:nzones_model_number_timestep_vs_age}
\end{figure}

We test temporal and spatial resolutions of our models by adjusting the following \texttt{MESA} settings: \texttt{time\_delta\_coeff}, which sets the temporal resolution, and \texttt{mesh\_delta\_coeff}, which sets spatial resolution. We lower these simultaneously in steps of 0.2 starting from 0.8 and ending at 0.4, increasing temporal and spatial resolutions with each step. In the panels of Fig. \ref{fig:nzones_model_number_timestep_vs_age} we illustrate how different quantities of our models are affected by lowering these settings. In all three panels, the coarsest model is drawn in purple, the model with finer resolution is drawn in orange, and the finest model is drawn in green. The quantities are drawn against "time since burst beginning", which corresponds the ages of the \texttt{burst\_model} in years, or "relative age" in the main text.

The top panel illustrates how the number of zones increases with lowering specifically \texttt{mesh\_delta\_coeff}: Between 0.8 and 0.4 the number of zones is effectively doubled, roughly from $4000$ to $8000$. The middle panel illustrates how many model numbers, that is, number of calculations, our models accumulate. Lowering \texttt{time\_delta\_coeff} demands shorter timesteps and consequently more calculations. The three models here accumulate respectively 13149, 15333, and 17624 model numbers. The bottom panel illustrates timestep sizes in years. Timesteps are naturally shorter for the finest model, as we demand finer temporal resolution. The dips in the timestep sizes are intentional and correlates with the output of structure profiles. During the burst we save profiles every ten years and post burst every fifty years. On the timescales that we consider, pulsation frequencies and structure properties can change quickly. It is therefore crucial that the ages at which profiles are outputted differ by no more that a few days between the models. Should \texttt{MESA} overshoot a target age for a profile save, a retry is demanded with a new timestep reduced by a factor $0.9$. This repeats until the age of the model matches the target age within some threshold that we set to $0.001 \ \text{years}$. This gradual lowering of the timestep sizes gives the periodic dips.

\subsection{Convergence of \texttt{MESA} calculations}
\begin{figure}[h!]
    \centering
    \includegraphics[width=1.0\linewidth]{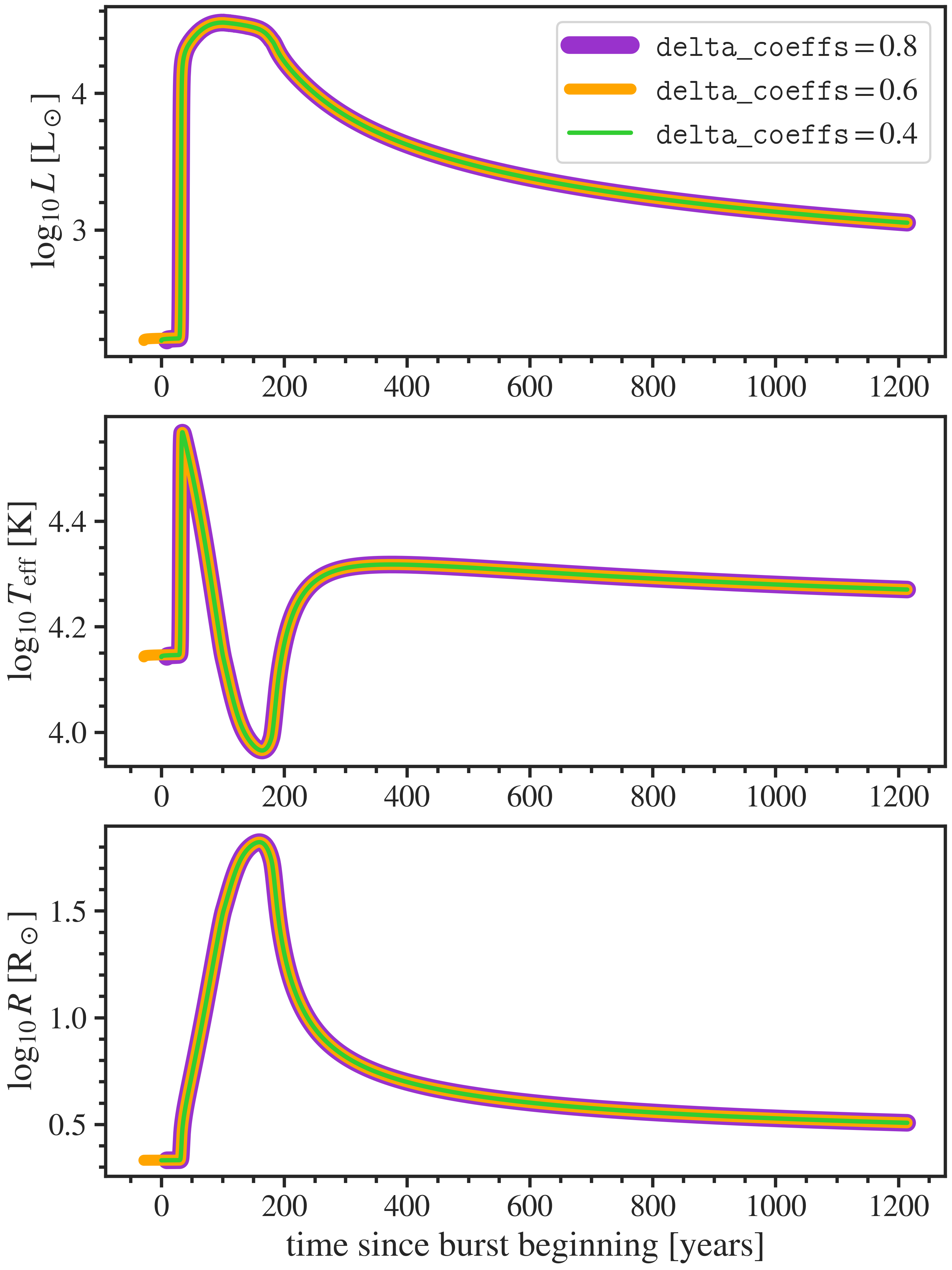}
    \caption{Model quantities drawn against time since burst beginning. The coarsest, finer, and finest models respectively have $\texttt{*\_delta\_coeff}$'s of 0.8, 0.6, and 0.4, and are drawn in purple, orange and green. Top panel illustrates surface luminosity, middle panel illustrates effective temperature, and bottom panel illustrates radius.}
    \label{fig:log_L_Teff_R_vs_age}
\end{figure}

We compare here properties of the three \texttt{MESA} models in Sec. \ref{Appendix A:Increasing resolution} for which temporal and spatial resolutions were gradually increased. Fig. \ref{fig:log_L_Teff_R_vs_age} compares global properties of the models across the burst. The coarsest, finer, and finest models are again drawn in purple, orange and green respectively. We draw in the top panel surface luminosity, in the middle panel effective temperature, and in the bottom panel radius. All quantities are drawn against time since burst beginning, as in Fig. \ref{fig:nzones_model_number_timestep_vs_age}.

Inspecting Fig. \ref{fig:log_L_Teff_R_vs_age}, we find good agreement between the models. The quantities drawn here mainly concern trajectories in the HR-diagram. Therefore, the burst-excursions we observe in the HR-diagram are insensitive to resolution changes.

\begin{figure}[h!]
    \centering
    \includegraphics[width=1.0\linewidth]{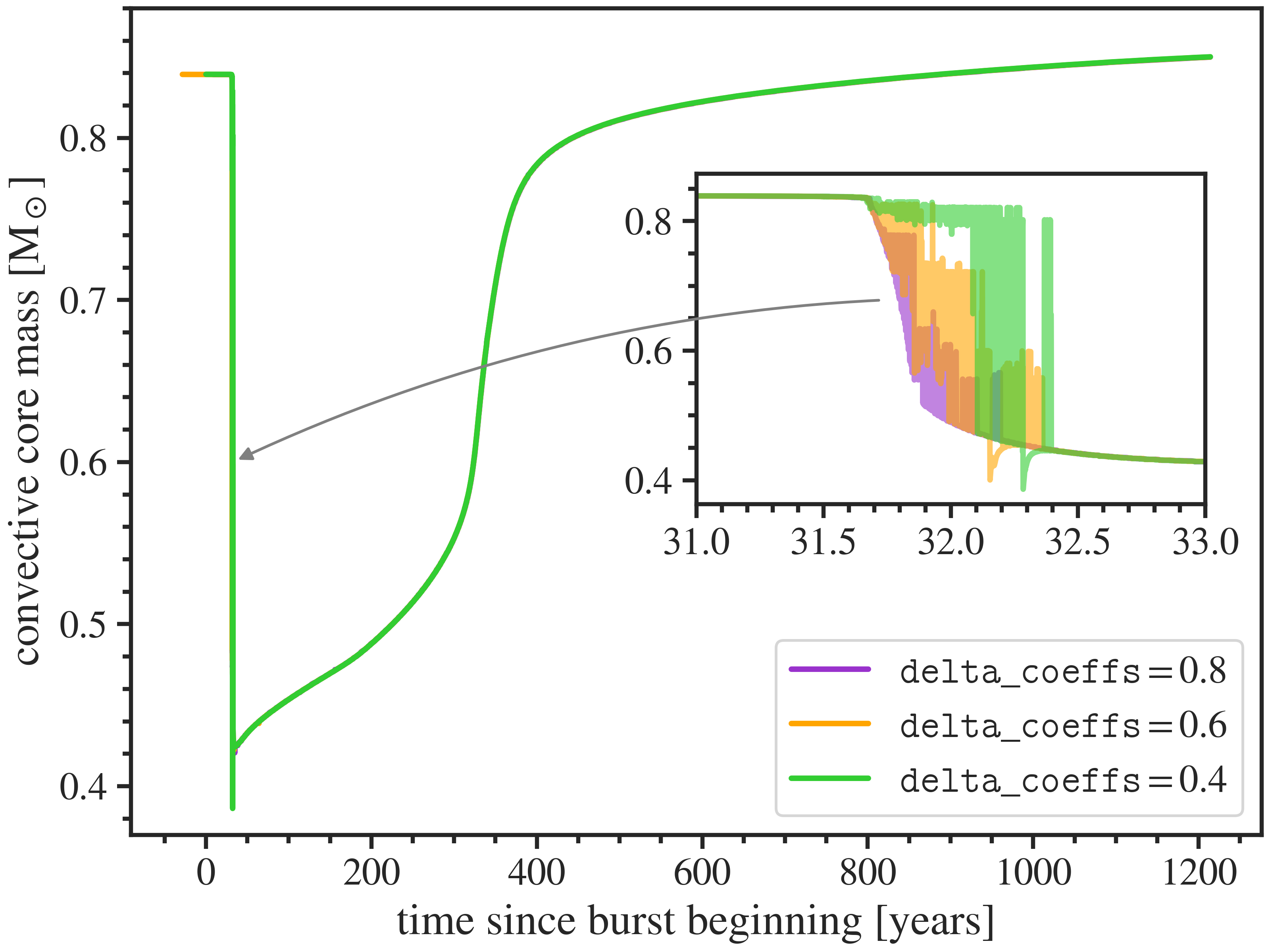}
    \caption{Convective core masses drawn against time since burst beginning. The coarsest, finer, and finest models respectively have $\texttt{*\_delta\_coeff}$'s of 0.8, 0.6, and 0.4, and are drawn in purple, orange and green. An inset figure highlights the roughly one year window in which core initially recedes.}
    \label{fig:convergence_convective_core_mass}
\end{figure}

A key component of our analysis is the receding convective core at the onsets of the accretion bursts. We therefore examined whether numerical convergence is attained for the convective core mass. This is illustrated in Fig. \ref{fig:convergence_convective_core_mass} where convective core masses of the three models are drawn against time since burst beginning. An inset figure shows a zoom-in on the window in which the core recedes.

Examining the outer diagram of Fig. \ref{fig:convergence_convective_core_mass}, we note good agreement between the models. On the longer timescales that this diagram represents, the evolutions of the convective core masses are indistinguishable across the different resolutions. The inset figure, however, tells a slightly different story. In the roughly one year window during which the core dramatically recedes our models do not agree. For example, the finest model (green) plateaus roughly for half a year before dropping to a core mass similar to the coarsest model (purple). As we show below, however, this discrepancy will not affect the conclusions of our work. Outside the window in which the core recedes the models once again find agreement. Similarly we are able to attain numerical convergence in the asteroseismic properties of our models.

\subsection{Convergence of \texttt{GYRE} calculations}
\begin{figure}[h!]
    \centering
    \includegraphics[width=1.0\linewidth]{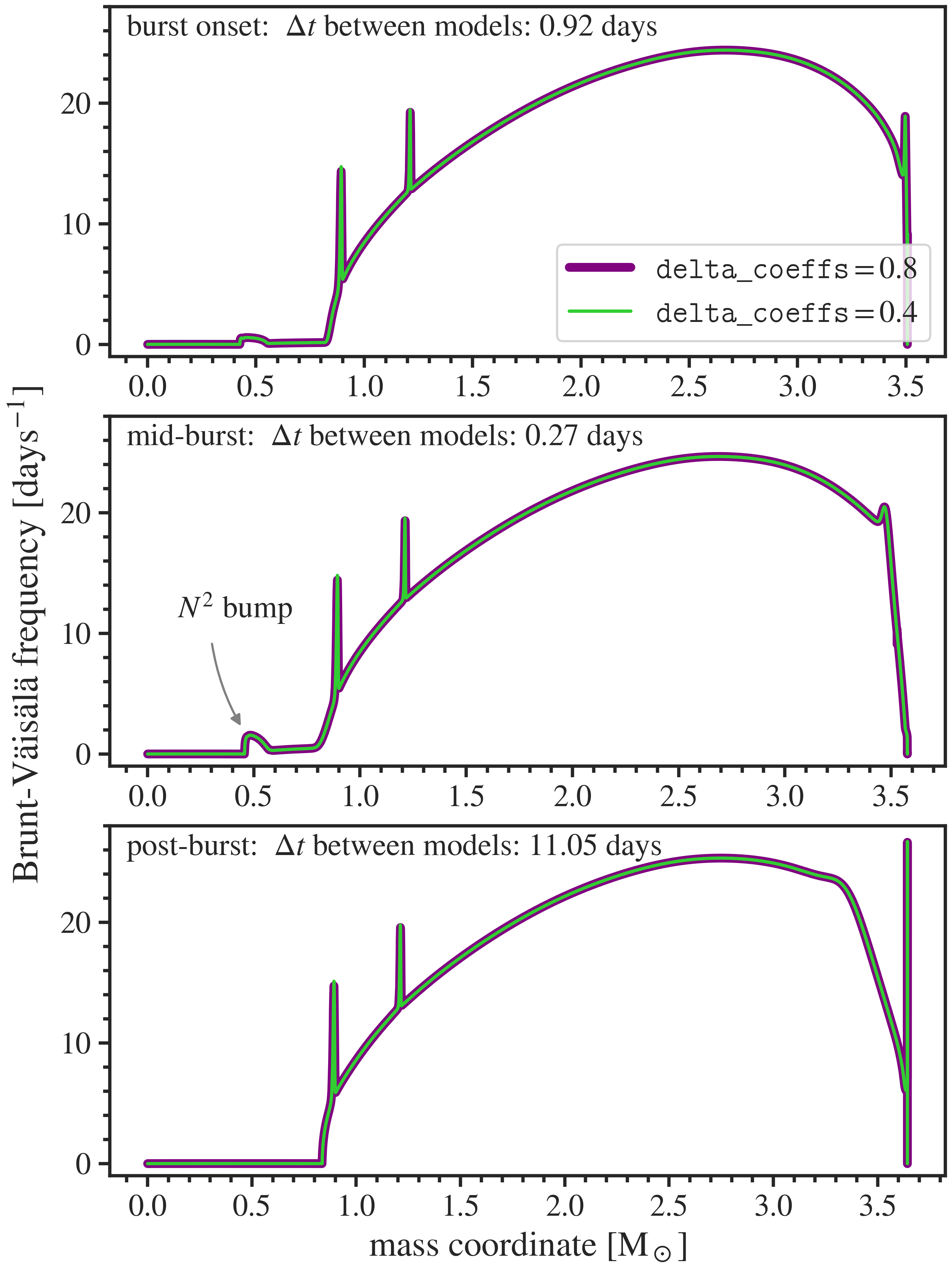}
    \caption{Brunt-Väisälä profiles against mass coordinate at different burst stages. The coarsest model is drawn in purple and the finest model is drawn in green. Top panel illustrates profiles at the onset of the burst, middle panel mid-burst, bottom panel post-burst.}
    \label{fig:convergence_brunt_N2}
\end{figure}

We consider now whether eigenfrequencies calculated with \texttt{GYRE} are insensitive to resolution changes in our \texttt{MESA} models. To this end, we first establish numerical convergence in the Brunt-Väisälä frequency profiles of our models. This is illustrated in Fig. \ref{fig:convergence_brunt_N2} where Brunt-Väisälä profiles for our coarsest (purple) and finest models (green), with \texttt{*\_delta\_coeff}'s 0.8 and 0.4 respectively, are drawn against mass coordinate. Top panel illustrates profiles at the onset of the burst, middle panel during the burst, and bottom panel post burst. Text in the top left corner also indicates the age differences between the models (see Sec. \ref{Appendix A:Increasing resolution}).

We observe good agreement in the Brunt-Väisälä profiles across the different resolutions and at different stages of the burst. The accretion bursts imprint themselves in the Brunt-Väisälä profiles as a bump directly above the convective core ($N^2$ bump at $0.5 \ \text{M}_\odot$ in middle panel). Between our coarsest and finest models the bumps are indistinguishable. This is important for the frequency calculations with \texttt{GYRE} as \textit{g}-modes propagate from the inner to the outer edge of the Brunt-Väisälä profiles. Had the bump not been consistent across different resolutions, our \texttt{GYRE} calculations would not have been consistent.

\begin{figure}[h!]
    \centering
    \includegraphics[width=1.0\linewidth]{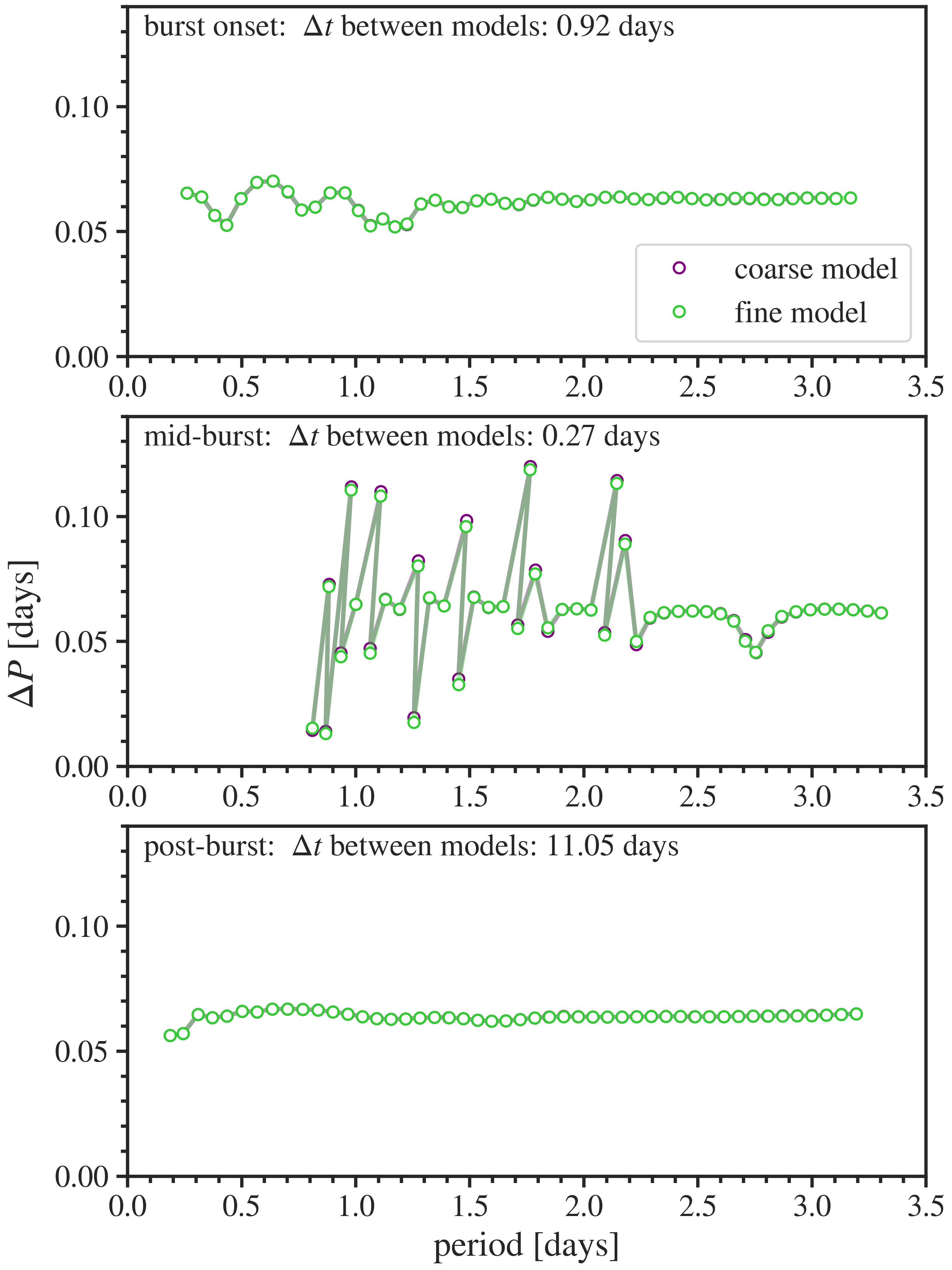}
    \caption{Period-spacing diagrams for the models of Fig. \ref{fig:convergence_brunt_N2} drawn in purple and green for the coarsest and finest model respectively.}
    \label{fig:convergence_period_spacing}
\end{figure}

Finally we assess numerical convergence in the frequency calculations with \texttt{GYRE}. This is showcased in Fig. \ref{fig:convergence_period_spacing} where period-spacings of the coarsest model are compared with spacings of the finest. The spacings are taken from the same ages as the panels in Fig. \ref{fig:convergence_brunt_N2} corresponding to burst onset, mid-burst, and post burst. In the middle panel we have used a slightly different plotting routine compared to the main text, hence the diagrams differ. The frequency-to-frequency comparison, however, is still valid.

We observe overall good agreement between the models during all burst stages considered here. While small differences are observed, for example, in the middle panel, these will not affect overall conclusions of our work.

% \section{Figures to support the main text}

% In this appendix we provide figures to support claims made in the main text.

\section{Burning figures}

\begin{figure}
    \centering
    \includegraphics[width=1.0\linewidth]{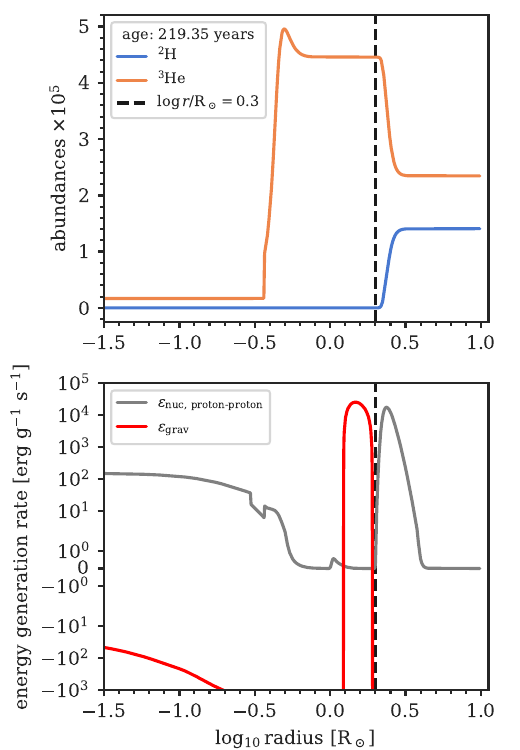}
    \caption{Abundance profiles for $\deuterium$ and $\hethree$ (top panel), and the profile of the energy generation rate from the proton-proton chain (bottom panel), illustrating that a pocket of positive $\egrav$ around $\log_{10}{r/\text{R}_\odot}\simeq0.3$ (see Fig. \ref{fig:eps_grav_heatmap}) coincides with deuterium burning.}
    \label{app:fig:deuterium}
\end{figure}

Fig. \ref{app:fig:deuterium} supports the claim in Sect. \ref{subsec:internal_structures} that the strip of positive $\egrav$ in Fig. \ref{fig:eps_grav_heatmap} coincides with deuterium burning from age 200 and onwards. In the top panel, the blue line is the deuterium abundance profile ($\deuterium$) at $\text{age}\approx220 \ \text{years}$, and the orange line is the profile of $\hethree$. The dashed black line is the location of the strip of positive $\egrav$. In the bottom panel the burning profile of the proton-proton (\textit{pp}) chain is drawn in gray and the profile of $\egrav$ is drawn in red, both on a symmetric logarithmic scale. 

Deuterium burning is evident: at $\log_{10}{r/\text{R}_\odot}\simeq0.3$ the abundance of deuterium drops to zero while $\hethree$ increases. At the same time, there is significant burning from the \textit{pp}-chain, and $\egrav$ climbs to positive values just below the burning spike. This follows from,
\begin{equation*}
    \frac{\partial L}{\partial m}-\epsilon_\text{nuc} < 0 \quad \Rightarrow \quad \egrav>0.
\end{equation*}
The sudden drop in $\epsilon_\text{nuc}$ below $\log_{10}{r/\text{R}_\odot}\simeq0.3$ means the release of gravothermal energy must become the main source of energy, that is, $\egrav>0$.

\begin{figure}
    \centering
    \includegraphics[width=1.0\linewidth]{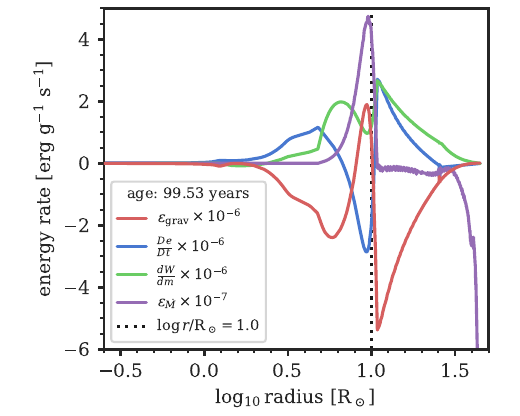}
    \caption{Profiles of energy generation rate components in Eq. (\ref{eq:egrav_paxton_2018}): $\egrav$ (red), $De/Dt$ (blue), $dW/dm$ (green), and $\epsilon_{\dot{M}}$ (purple). This figure illustrates that the pocket of positive $\egrav$ around $\log_{10}{r/\text{R}_\odot}\simeq1.0$ in Fig. \ref{fig:eps_grav_heatmap} originates from a decrease in the differential work and the internal energy at this location.}
    \label{app:fig:components}
\end{figure}

Fig. \ref{app:fig:components} illustrates how different components of the luminosity gradient behave during the pocket of strong $\egrav$ located at $\log_{10}{r/\text{R}_\odot}\simeq1.0$ (Fig. \ref{fig:eps_grav_heatmap}). The components are $\egrav$ (red), $De/Dt$ (blue), differential work ($dW/dm$; green), and $\epsilon_{\dot{M}}$ (purple).

At $\log_{10}{r/\text{R}_\odot}\simeq1.0$ both $De/Dt$ and $dW/dm$ decrease, leading to an increase in $\egrav$. Simultaneously, there is a significant increase in $\epsilon_{\dot{M}}$ (note an order of magnitude difference in the scaling between $\epsilon_{\dot{M}}$ and the other components).

\section{Brunt-Väisälä figures}

\subsection{Full Brunt-Väisälä profiles to explain the drop in the fundamental spacing}

\begin{figure}
    \centering
    \includegraphics[width=1.0\linewidth]{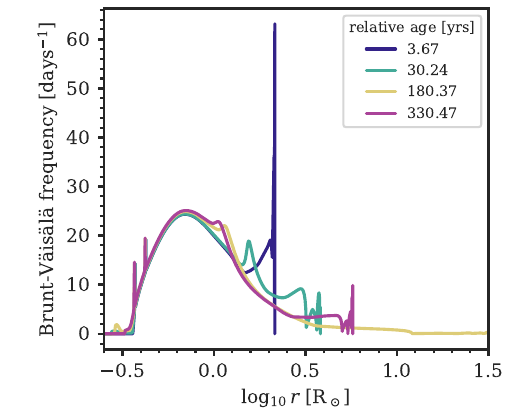}
    \caption{Full Brunt-Väisälä profiles drawn against radius. We refer to this figure to explain the changes in $\Pi_0$ during the burst.}
    \label{app:fig:brunt_N_evolution}
\end{figure}

Fig. \ref{app:fig:brunt_N_evolution} illustrates how the shape of the Brunt-Väisälä profile changes across the burst for the model undergoing $\dot{M}_\text{burst}=10^{-3} \ \mdot$. This, in part, explains the drop in $\Pi_0$, however, the expansion of the star also contributes.

\subsection{Zooming in on the outermost layers}
\begin{figure}
    \centering
    \includegraphics[width=1.0\linewidth]{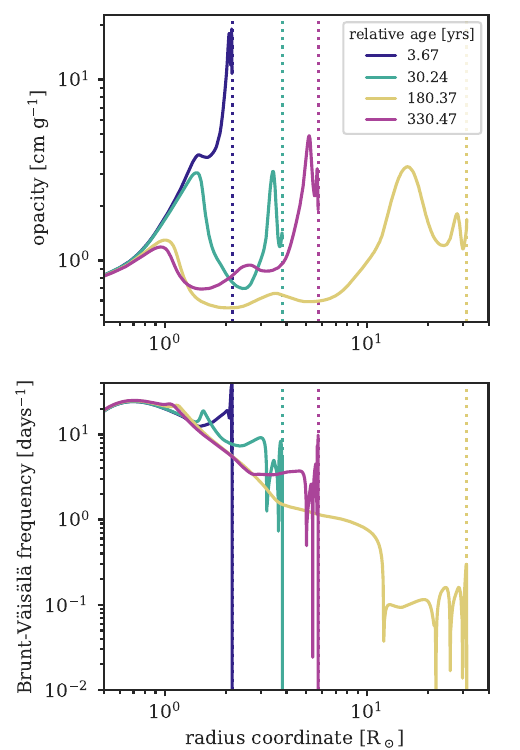}
    \caption{Bottom panel shows the same Brunt-Väisälä profiles as Fig. \ref{fig:chemistry}, but zoomed in out the outer layers of the star rather than the convective-core boundary. In the top panel, the corresponding opacity profiles are drawn.}
    \label{fig:app:BV_surf}
\end{figure}

The same models as in Fig. \ref{fig:chemistry} are drawn in Fig. \ref{fig:app:BV_surf}, except this figure showcases only the surface regions of the models. Bottom panels shows the Brunt-Väisälä profiles ($N$) and in the top panel opacity profiles are drawn. The surface is indicated by dashed colored lines, and both $N$, opacity, and radius are drawn on logarithmic scales to highlight the features of interest.

% \begin{figure}
%     \centering
%     \includegraphics[width=1.0\linewidth]{updated_figures/BV_surf.pdf}
%     \caption{Brunt-Väisälä profiles for the same models as in Fig. \ref{fig:chemistry} (blue, green, yellow, and purple lines). Whereas Fig. \ref{fig:chemistry} showcased the imprint in the core, this figure shows how the Brunt-Väisälä profiles change near the surface. Dashed lines indicate the surface coordinate; note that it moves because the star expands.}
%     \label{fig:BV_surf}
% \end{figure}

Fig. \ref{fig:app:BV_surf} reveals that $N$ stretches outward as the star expands. In model A (dark blue) the turning point of the \textit{g}-modes is effectively vertical. In contrast, for models B, C and D (green, yellow and purple lines), the outer edge of $N$ has irregularities. Some ambiguity exists in determining whether the frequency shifts are created from the core or the surface regions \citep{Montgomery2003}. Consequently, the calculated frequencies of our models (Sect. \ref{sec:PSP}) may also be sensitive to the shapes of $N$ near the surface. We find that the shapes of $N$ here cannot be pinpointed to chemical gradients, but find that the features coincide with changes in the opacity profiles, which in turn feed into the temperature gradients ($\nabla_T$) and affect the shapes of $N$ in each model (Eq. \ref{eq:brunt_N2}).

\section{Iterative smoothing of the FT of the period-spacing pattern}

\begin{figure}
    \centering
    \includegraphics[width=1.0\linewidth]{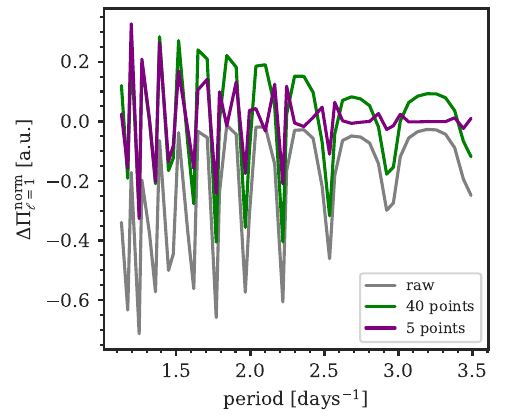}
    \caption{Period-spacing pattern of Fig. \ref{fig:period_spacing_result} with iterative smoothing by Savitzky-Golay filters (to support Sect. \ref{subsec:FT_of_PSP}).}
    \label{fig:app:smoothed_psp}
\end{figure}

Fig. \ref{fig:app:smoothed_psp} shows the iteratively smoothed period-spacing pattern discussed in Sect. \ref{subsec:FT_of_PSP}.

\section{Rotation}\label{app:rotation}
Young intermediate-mass stars (YIMS; B-types included) form in the gravitational collapse of magnetized, high-density, rotating, and turbulent cloud fragments. Because the original cloud has net angular moment, infalling matter flattens into a disk aligned along the orbital plane of the central protostar \citep{Hartmann2009}. The ensuing accretion from disk to protostar transfers both mass and angular momentum, resulting in spin-up of the protostar. Combined with the spin-up from pre-MS contraction, rough estimates of the rotation rates of YIMS are near critical \citep[see, for example,][]{Rosen2012}. Observational evidence, however, points to spin-down mechanisms, with studies reporting the following fractions \citep{Huang2010} : $6\%$ slow rotators ($V_\text{eq}/V_\text{crit}<0.1$), $52\%$ with values $0.4<V_\text{eq}/V_\text{crit}<0.8$, and $1.3\%$ fast rotators ($V_\text{eq}/V_\text{crit}>0.9$) in a sample of 220 young B-type stars. Here $V_\text{eq}$ is the equatorial rotational velocity, and $V_\text{crit}$ is the critical break-up velocity. Proposed spin-down mechanisms include the magnetic torque \citep{Matt2005, Matt2010, Rosen2012}, gravitational torque in the earliest stages of protostar formation \citep{Lin2011}, and angular momentum removal by accretion-powered stellar winds \citep{Matt2005:winds}. In the case of SPB stars, \textit{g}-modes may also transport angular momentum across the stellar interior \citep{GYRE2018}.

The rotational velocities highlighted above are sufficient to shift the \textit{g}-mode periods, pointing to rotation playing a bigger role during the protostellar- and pre-MS phases than explored in this work. While a full analysis on rotation is out of the scope of this work, we perform here \texttt{GYRE} computations with rotation to illustrate its effects on the predicted period-spacing patterns. We restrict our inclusion of rotation solely to \texttt{GYRE}'s non-perturbative treatment of the Coriolis force based on the traditional approximation of rotation \citep[TAR;][]{eckart1960hydrodynamics, Townsend2020:rot} with uniform rotation. In the co-rotating frame of reference, the TAR leads to a \textit{g}-mode period-spacing pattern determined by \citep[see, for example,][]{Aerts2021},
\begin{equation}\label{app:eq:TAR}
    \Delta \Pi^\text{co}_{\ell m s}=\frac{\Pi_0}{\sqrt{\lambda_{\ell m s}}},
\end{equation}
where $\lambda_{\ell m s}$ is the eigenvalue for the \textit{g}-mode of spherical degree $\ell$ and azimuthal order $m$, obtained from solving the Laplace tidal equation in a star with spin parameter $s=2\tilde{\Omega}/\omega$, where $\tilde{\Omega}$ is the angular velocity of the star and $\omega$ is the angular mode frequency in the co-rotating frame \citep[see, for example,][]{Mathis2019TAR}. Conversely, in the inertial frame, an additional shift is expected from the Doppler effect. Nevertheless, we restrict ourselves to $m=0$ modes for which the frequency shifts in the inertial- and co-rotating frames are identical \citep[][]{Ouazzani2017,Henneco2024}.

\begin{figure}
    \centering
    \includegraphics[width=1.0\linewidth]{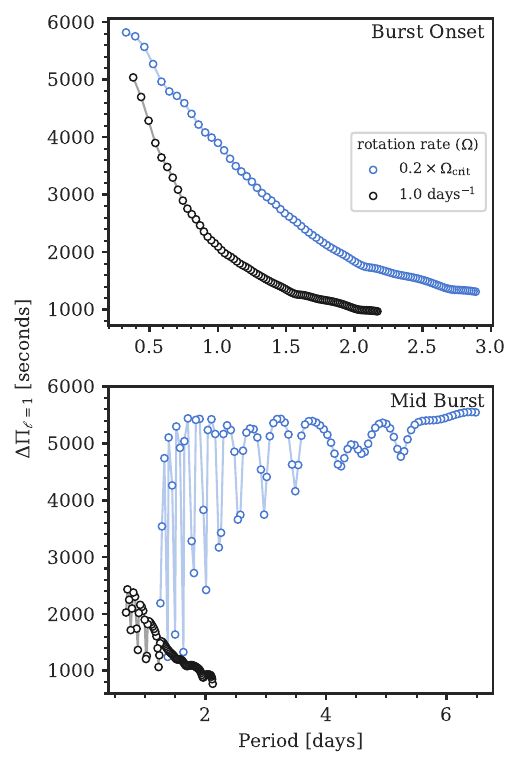}
    \caption{Period-spacing patterns for the modes $(\ell,m)=(1,0)$ for two different implementations of uniform rotation with \texttt{GYRE}. The model in black is computed with a constant rotation frequency of $1.0 \ \text{d}^{-1}$ and the blue model with an angular rotation frequency fixed to $20\%$ of the critical ($\Omega_\text{crit}$). Top panel shows the period-spacings moments before the burst is initiated, and the bottom panel captures them at an arbitrary moment during the burst.}
    \label{app:fig:PSP_rot}
\end{figure}

We illustrate the effects of rotation in the inertial frame for two cases. First, we employ the TAR with \texttt{GYRE} for a constant rotation frequency of $\nu_\text{rot}=1.0 \ \text{d}^{-1}$ across the entire accretion burst (R1). Second, we fix the angular rotation frequency to $20\%$ of the critical angular frequency ($\Omega_\text{crit}$; R2). The resulting period-spacing diagrams for the $(\ell,m)=(1,0)$ modes are shown in Fig. \ref{app:fig:PSP_rot}, where R1 is drawn in black and R2 in blue. The top panel shows the period-spacings just before the burst is initiated, while the bottom panel shows them during the burst. In the simplest approximation, the critical rotation is marked by the balance of the gravitational acceleration with the centrifugal at the equator ($g_\text{eq}=a_\text{eq}^\text{cf}\Leftrightarrow GM/R_\text{eq}^2=\Omega^2_\text{crit} R_\text{eq}$) resulting in the following expression $\Omega_\text{crit} [\text{rad}~\text{s}^{-1}]=\sqrt{GM/R_\text{eq}^3}$. Converting to frequencies in units $\text{d}^{-1}$ ($\nu_\text{crit}=\Omega_\text{crit} \times86400~\text{s}~\text{d}^{-1}/2\pi$) we get $\nu_\text{crit}=5.113 \ \text{d}^{-1}$ and $0.033 \ \text{d}^{-1}$ for R2 in the top- and bottom panels of Fig. \ref{app:fig:PSP_rot} respectively. The actual rotation frequencies become $\nu_\text{rot}=0.2\nu_\text{crit}=1.02 \ \text{d}^{-1}$ and $0.007 \ \text{d}^{-1}$. We may conclude from the derivation above that the drop in rotation frequency is due to the rapid expansion of the star during the burst and conservation of angular momentum \citep[these values can shift with the inclusion of more physical effects, see, for example,][]{Lamers2017,GYRE2018}.

In the top panel, the period-spacing patterns of both R1 and R2 exhibit the slopes expected from Eq. (\ref{app:eq:TAR}). Furthermore, in the R2 top-panel configuration, spin parameters are in the range $10<s<190$, meaning all the modes are sub-inertial \citep[if not strongly sub-inertial; ][]{Aerts2024:rotation}. In this case, the Coriolis force is not a simple perturbation, the effects of which are seen in the steep slopes of the period-spacing patterns.

In the bottom panel, R1 retains the slope but now also shows multiple dips, originating from the imprint discussed in the main text. The approximate $\nu_\text{crit}$ derived above for the mid-burst stage is nearly two orders of magnitude smaller than the $\nu=1.0 \ \text{d}^{-1}$ explored here. Therefore, the configuration presented for R1 in the bottom panel is likely unphysical. R2, however, has returned to a period-spacing with no slope, but the dips are still present. In the TAR, the no-slope feature is recovered for spin parameter $s\to0$ (we find $s<0.02$), or, in other words, slow rotation \citep{Aerts2021}. The flattening of the period-spacing for R2 is explained by the radial inflation and spin-down of the star. As a result, the Coriolis force plays an increasingly smaller role in acting as restoring force, and the perturbation to the frequencies is smaller.

A final effect we did not consider in this work is the coupling of pure inertial waves in the convective core with gravito-inertial waves in the envelope. The signature of this coupling is a distinct dip-like structure in the period-spacing diagram, sensitive to the near-core boundary \citep{Ouazzani2019, Tokuno2022, Barrault2025, Breton2026}. Given our asteroseismic imprint also appears at this boundary, we may suspect an interplay between the two effects. We defer to this as the subject of future work.

The actual rotational evolution of a protostar is governed by a variety of physical mechanisms. In addition to the aforementioned star-disk interactions, rotation is also sensitive to the characteristics of the disk and the accretion process itself. For example, the inclusion of hot accretion generally results in more bloated objects, affecting the star-disk interaction, even during quiescent accretion \citep{Gehrig2023}. As no detailed study exists on the rotational evolution of a protostar during an FUOr-type accretion burst, we can only speculate on its impact. Significant amounts of angular momentum will be transferred to the protostar, resulting in spin-up. This spin-up, to some extent, can be assumed counteracted by the radial inflation and strong mass transfer during the burst. Given that the amount of accretion energy injected into the star is assumed to increase with the accretion rate \citep{Steindl2022, Joerjensen2025}, we may also speculate that the results presented in this work underestimate the actual bloating a protostar undergoes during an FUOr burst, as we have not included hot accretion.

\end{appendix}

\end{document}